%% file: main.tex
\documentclass[a4paper, 12pt, reqno, tbtags, final]{amsart}
\usepackage{amsmath, amsthm, amssymb, amsrefs, bbm}

\usepackage{verbatim}
\usepackage{fullpage}
\usepackage{enumerate}
\usepackage{fancybox}
\usepackage{cases}
\usepackage{graphicx}
\usepackage{hhline}
\usepackage{accents}

\input{defs.tex}

\title{Long-Term Mean-Variance Optimization Under Mean-Reverting Equity Returns}
\author{Michael Preisel}
\date{6 March 2024}
\address{Universitetsparken 5, 2100 Copenhagen E, Denmark}
\email{mipr@math.ku.dk}
\urladdr{www.math.ku.dk}

\begin{document}
\input{abstract.tex}
\maketitle
\bigskip
\bigskip
\input{introduction.tex}
\input{capmktmodel.tex}
\input{portfoliodyn.tex}

\input{optimnom.tex}

\input{spectral_problem.tex}
\input{general_solution.tex}
\input{discussion.tex}
\input{conclusion.tex}

\appendix
\input{capmkt_appendix.tex}

\input{prop_xidef.tex}
\input{prop_nominal_meanvar.tex}
\input{thm_nominal_euler.tex}
\input{cor_latent_roots.tex}
\input{thm_nominal_solution.tex}
\input{lemma_meanvar.tex}
\input{cor_infiniteriskaversion.tex}

\bibliography{main.bbl}

\end{document}

%% file: defs.tex
\newcommand{\beq}{\begin{equation}}
\newcommand{\eeq}{\end{equation}}
\newcommand{\vasicek}{{Va\v{s}\'{i}\v{c}ek}}

\newtheorem{thm}{Theorem}[section]
\newtheorem{prop}[thm]{Proposition}
\newtheorem{lem}[thm]{Lemma}
\newtheorem{cor}[thm]{Corollary}

\DeclareMathOperator{\R}{\mathbb{R}}
\DeclareMathOperator{\E}{\mathbb{E}}
\DeclareMathOperator{\V}{\mathbb{V}}
\DeclareMathOperator{\C}{\mathbb{C}}

\newcommand{\unitI}{\mathbb{I}}
\newcommand{\boldZero}{\boldsymbol{0}}
\newcommand{\rbar}{\bar{r}}
\newcommand{\xbar}{\bar{x}}

\newcommand{\xibar}{\bar{\xi}}

\newcommand{\boldA}{\boldsymbol{A}}
\newcommand{\boldB}{\boldsymbol{B}}
\newcommand{\boldC}{\boldsymbol{C}}
\newcommand{\boldD}{\boldsymbol{D}}

\newcommand{\boldG}{\boldsymbol{G}}
\newcommand{\boldH}{\boldsymbol{H}}

\newcommand{\boldK}{\boldsymbol{K}}

\newcommand{\boldM}{\boldsymbol{M}}
\newcommand{\boldS}{\boldsymbol{S}}

\newcommand{\boldQ}{\boldsymbol{Q}}

\newcommand{\boldGamma}{\boldsymbol{\Gamma}}

\newcommand{\boldPsi}{\boldsymbol{\Psi}}
\newcommand{\boldLambda}{\boldsymbol{\Lambda}}

\newcommand{\boldxi}{\boldsymbol{\xi}}

%% file: abstract.tex
\begin{abstract}
This paper studies the mean-variance optimal portfolio choice of an investor pre-committed to a deterministic investment policy in continuous time in a market with mean-reversion in the risk-free rate and the equity risk-premium.

In the tradition of Markowitz, optimal policies are restricted to a subclass of factor exposures in which losses cannot exceed initial capital and it is shown that the optimal policy is characterized by an Euler-Lagrange equation derived by the method of Calculus of Variations. 

It is a main result, that the Euler-Lagrange equation can be recast into a matrix differential equation by an integral transformation of the factor exposure and that the solution to the characteristic equation can be parametrized by the eigenvalues of the associated lambda-matrix, hence, the optimization problem is equivalent to a spectral problem.

Finally, explicit solutions to the optimal policy are provided by application of suitable boundary conditions and it is demonstrated that - if in fact the equity risk-premium is slowly mean-reverting - then investors committing to long investment horizons realize better risk-return trade-offs than investors with shorter investment horizons.

 \newline \noindent  \newline \noindent
This work is sponsored by Railpen, 7th Floor, 100 Liverpool St, London EC2M 2AT, United Kingdom.
\bigskip

\end{abstract}

%% file: introduction.tex
\section{Introduction}
It is a long-standing investment advice that younger people should hold higher proportions of risky assets than other investors given their longer investment horizon. This implies that - on a risk-adjusted basis - it is rewarded simply to hold a risky assets for an extended period of time. It is therefore a statement of - or an assumption about - capital market structure which is supported by the literature on rational portfolio choice only if returns on risky assets display non-trivial behaviour, say, if returns are mean reverting \cites{samuelson89, kim96, campbell99, barberis00, wachter02, munk04}. 

Another aspect is, if young people actually would hold risky positions over sustained periods of time. One such motive would be to save for retirement. Retirement savings are often organized through specialized financial institutions tasked with converting funds put aside during working life into an income stream during retirement. Common to such pension providers\footnote{The term 'pension provider'  covers any savings-based pension arrangement, say, pension funds, individual accounts, target date funds, etc.} is, that funds committed in working life will not be released until after retirement, hence, the savings vehicle institutionalizes long holding periods.  Therefore, pension providers are - on behalf of younger people - examples of 'long-term' investors in the sense of this paper as they truly can plan capital allocation and (buy and) hold assets over very long periods of time. 

This of course raises the question of what investment strategy a pension provider should apply. By tradition, for reasons of governance or - often - simply by regulation, the current and future portfolio composition is set by a strategic asset allocation, a deterministic investment strategy detailing the allocation to major asset classes, say, stocks and bonds, from present day to all future dates. Since the pension provider - at least formally - pre-commits to a fixed investment strategy the optimization problem is reduced to an optimization over deterministic investment strategies only - see \cites{christiansen2013b, christiansen2018} for similar applications to retirement savings.

Since Markowitz \cite{markowitz52}, the workhorse of the asset management industry remains mean-variance optimization. The original work was a single-period optimization but extension to multiple periods has proven not to be straightforward and the literature on this topic is vast, see \cites{steinbach01,schweizer10} and references therein.

In combination, a long-term investor therefore would seek the (deterministic) investment strategy to maximize the expected return on a given horizon for a fixed variance target. If asset-returns mean-revert over time-scales comparable to  or longer than  the investment horizon then - in support of the claim to sustain larger allocations to risky assets - the risk-return trade-off should improve with investment horizon as a testament to long-term risk being rewarded.

In the case of continuous-time mean-variance optimization, it is well known that the optimization problem cannot be solved by standard dynamic programming techniques. Several authors have addressed this problem: In \cite{cox89}, it is shown that for complete markets the distribution of optimal wealth can be derived by martingale methods, hence, the optimal strategi is unique and exists; \cite{basak10} extends solutions to incomplete markets by splitting the optimization criterion in to a local term and a term taking time-inconsistency into account to apply a dynamic program; \cite{steinbach01} provides an exact recursive solution for a multi-period portfolio tree; \cite{bjork09} proposes a game-theoretic approach by which the portfolio is optimized recursively taking into account each subsequent allocations will do the same; \cite{czichowsky13} proposes the concept of local mean-variance efficient portfolios by which the portfolio remains mean-variance efficient over any subperiod, and \cites{christiansen2013a, christiansen2013b} derives a characterizing set of differential equations from the Hamilton-Jacobi-Bellman and Pontryagin minimization principles for deterministic investment strategies for a specialized version of the optimization criterium. 

Common to these approaches is that analytical solutions only are derived for Black-Scholes type of capital markets - otherwise only numerical solutions are provided.

The objective of this paper therefore is - for a given investment horizon - to derive explicit optimal investment strategies maximizing horizon expected return for fixed horizon variance in a capital market providing a non-trivial investment opportunity set in bonds and stocks, respectively, where bond returns follow a \vasicek\  model and stock returns display mean-reversion in surplus return.  Consistent with the practice of Strategic Asset Allocation, investment strategies are allowed to change deterministically over time but not to depend on the state of the capital market.

The capital market model was first proposed by \cite{munk04} where it provides the opportunity set to an investor optimizing a real CRRA utility on a fixed horizon but limited to investing in stocks and nominal bonds only. The capital market model was later extended to include index-linked bonds in \cite{jp17} where also an algorithm for exact simulation is derived. For the purposes of this paper, the model is reduced to nominal assets only.

The solution of the mean-variance optimization problem follows \cite{jarner21} who recently demonstrated by direct methods that explicit solutions can be derived from the principle of Calculus of Variations in the special case of zero correlation between stocks and bonds. 

The main result of this paper is to show that by an integral transformation of the factor allocations to cash, bonds, and equity, the Euler-Lagrange\ equation of the full optimization problem is an inhomogegeous second-order matrix differential equation in the transformed variable. Furthermore, from the theory of solvents \cite{dennis73} it is shown that the homogeneous solution is equivalent to a spectral problem and explicit formulas for the eigenvalues are provided. Finally, by careful analysis of boundary conditions explicit solutions to the optimization problem is provided.

%% file: capmktmodel.tex
\section{Capital Market Model} 
We assume the capital market provides investment opportunity in one risky asset, $S_u$, at time $u$ which - in line with common terminology - we will refer to as 'equity'. 

Consistent with the assumption of risk being rewarded long-term, the equity surplus-return, $x_u$, displays mean-reversion and is (perfectly) negatively correlated with equity returns, that is, equity returns will tend to be higher following a loss and lower following a gain. This allows the model to distinguish between short-term fluctuations and longer-term volatility which will be suppressed from the mean-reversion of the equity surplus-return, hence, holding equity is explicitly rewarded on long investment horizons - see also the discussion of model properties in \cite{jp17}.

To allow full control of horizon variance, the capital market model also explicitly provides investment options in an equilibrium bond market by providing a full nominal term structures of interest rates at all points in time as discussed in Section \ref{sec:bonds} below.

\subsection{Three Factor Capital Market Model}
The capital market model is stated by its real-world dynamics - or P-dynamics - reflecting the risks a holder of securities in the market faces. The equity return, $S_u$, satisfies the stochastic differential equation
\begin{equation}\label{dS_t}
	\frac{dS_u}{S_u} = (r_u + x_u)du + \sigma_S dW^S_u
\end{equation}
at time $u$ where $r_u$ is the (nominal) short rate, $x_u$ a state-dependent equity surplus-return, $\sigma_S$ is equity volatility, and $W^S_u$ is a standard Brownian motion.

The equity surplus-return is mean-reverting and assumed to follow an Ornstein-Uhlen\-beck process given by
\begin{equation}
\label{dx_t}
	dx_u = \alpha(\xbar - x_u) du - \sigma_x dW^S_u
\end{equation}
with $\alpha$ the mean-reversion strength, $\xbar$ the average surplus-return, and $\sigma_x$ the volatility of the equity surplus-return. The equity surplus-return is assumed to be perfectly negatively correlated with equity returns, hence, it loads on the same Brownian motion driving the stock index, $W^S_u$, albeit with a negative sign.

The short rate, $r_u$, is also assumed to follow an Ornstein-Uhlenbeck process and is given by
\begin{equation}
\label{dr_t}
	dr_u = \kappa (\rbar - r_u) du + \sigma_r dW^r_u
\end{equation}
where $\kappa$ is the mean-reversion strength, $\rbar$ is the (long-term) average interest rate, $\sigma_r$ is the interest rate volatility, and $W^r_u$ is a standard Brownian motion.

In total, the capital market model defines three state variables driven by a two-dimensional Brownian motion,
\begin{equation*}\label{Brownian_motions}
	W_u = (W^r_u, W^S_u)^T,
\end{equation*}
with correlation $\rho$. We will further assume the model is not degenerate, that is, $\rho^2<1$. The key features of the model are summarized in Appendix \ref{appendix_capmkt} where also explicit solutions for the state variables can be found.

\subsection{Pricing Consistency and the Risk Premium}
It is well known from arbitrage theory \cite{bjork04}, that in a complete market, the risk premium, $\xi_u$, is the stochastic process under which the discounted price process, $p_u/b_u$, becomes a martingale under a new risk-neutral measure, $Q$, defined by the Girsanov transformation
\begin{equation*}
	dW_u = dW^Q_u - \xi_u du 
\end{equation*}
where $p_u$ is the price process of any security in the market, $W^Q_u$ is a Brownian motion under $Q$, and $b_s = \exp (\int_t^s r_u du )$ is the bank account numeraire at time $s$ given initial time $t$. Hence, the risk-neutral - or Q - dynamics of assets is
\begin{subequations}
	\begin{align} \label{drQ}
		dr_u & = \big[\kappa (\rbar - r_u) - \sigma_r \xi^r_u \big] du + \sigma_r dW^{Q(r)}_u , \\
		\label{dSQ} \frac{dS_u}{S_u} & = (r_u + x_u - \sigma_S \xi^S_u)du + \sigma_S dW^{Q(S)}_u .
	\end{align}
\end{subequations}

Upon inspection of \eqref{dSQ}, it is immediately clear that $x_u/\sigma_S$ is the equity risk-premium. Furthermore, it is easily verified that if the risk-premium process is parameterized as
\begin{equation} \label{xi_t}
	\xi_u = (\xi^r_u, \xi^S_u)^T = 
	\begin{pmatrix}
		[(a-\kappa) r_u + \kappa \rbar - ab]/\sigma_r \\
		x_u/\sigma_S
	\end{pmatrix}
\end{equation}
where $\xi^r_u$ and $\xi^S_u$ are the interest rate and equity risk-premium, respectively, then the short rate, $dr_u = a(b-r_u) + \sigma_r dW^{Q(r)}_u$, retains its structure as an Ornstein-Uhlenbeck process under Q-measure, hence, it is a \vasicek\ model and the term structure of interest rates is well known \cite{vasicek77}.

For future reference, we state the capital market assumptions in the following proposition:
\begin{prop} \label{prop_xi_definition}
	Assume that volatilies $\sigma_r, \sigma_x, \sigma_S > 0$, mean-reversion strengths $\kappa \neq \alpha \neq a >0$, mean levels $\rbar, \xbar, b$ are scalar constants, and $\rho^2<1$ then for $s > t$ the risk-premium process, $\xi_s$, can be decomposed into
	\begin{equation} \label{riskpremium_definition}
		\xi_s = \xibar +  e^{-\boldGamma (s-t)}\xi_t +   \boldxi\,\int_t^s  e^{-\boldGamma (s-u)} dW_u
	\end{equation}
	with respect to initial values $r_t, x_t$ at time $t$ where
	\begin{alignat}{4} \nonumber
		&\xibar && = \begin{pmatrix} a(\rbar - b)/\sigma_r \\ \xbar/\sigma_S \end{pmatrix}, 
		&\xi_t  & = \begin{pmatrix} (r_t - \rbar)(a-\kappa)/\sigma_r \\ (x_t - \xbar)/\sigma_S  \end{pmatrix}, \\
		\intertext{and}
		%\label{lambda-definition}
		& \boldxi && = \begin{Bmatrix} (a-\kappa) & 0 \\ 0 & -\sigma_x/\sigma_S \end{Bmatrix}, \qquad
		%\label{boldGamma-definition}
		&\boldGamma & = \begin{Bmatrix} \kappa & 0 \\ 0 & \alpha \end{Bmatrix}   \nonumber
	\end{alignat}
	are diagonal matrices of full rank.
	
	Furthermore, the risk-premium, $\xi_s$, is Normally distributed with conditional mean
	\begin{equation*}%\label{mean_lambda_s}
		\E \xi_{s|t} = \xibar + e^{-\boldGamma (s-t)} \xi_t
	\end{equation*}
	and conditional variance
	\begin{equation*}%\label{variance_xi_s}
		\V \xi_{s|t} = \boldxi \boldsymbol{V}^\xi_{s|t} \boldxi
	\end{equation*}
	where
	\begin{equation}\label{lambda_V_matrix}
		\boldsymbol{V}^\xi_{s|t} = \begin{Bmatrix}
			\psi_{2\kappa}(s-t) & \rho\psi_{\kappa+\alpha}(s-t) \\
			\rho\psi_{\kappa+\alpha}(s-t) & \psi_{2\alpha}(s-t)
		\end{Bmatrix}
	\end{equation}
	and
	\begin{equation} \label{psi}
		\psi_\alpha(s-t) = \int_t^s e^{-\alpha (u-t)} du = \frac{1}{\alpha} \big(1-e^{-\alpha(s-t)} \big).
	\end{equation}

	\begin{proof}
		See appendix \ref{proof_prop_xi_definition}.
	\end{proof}
\end{prop}

It is clear from \eqref{riskpremium_definition} that the local (short-term) expected risk-premium is stochastic and in general deviates from the average risk-premium. Similarly, from \eqref{lambda_V_matrix} and the properties of \eqref{psi}, local variance initially grows linearly with investment horizon, $s$,  whereas 'in the long-run', the risk-premium is stationary with asymptotic variance, \eqref{asymptotic_variance_xi_s}, and expected return $\xibar$, hence, the model explicitly allows 'short-term' risk-premium characteristics to be distinguishable from 'long-term'.

\subsubsection{Bonds}\label{sec:bonds}
As discussed above, the interest-rate risk-premium process under $Q$-measure is a \vasicek\ model, hence, in addition to the risky asset, the model provides an equilibrium bond-market consistently priced by the \vasicek\ pricing formula. 

To simplify the exposition, we will - similar to the equity index - assume the existence of a single zero-coupon bond, $B_u$, with fixed maturity, $M_B$, coinciding with - or beyond - the investment horizon, $s$. Since the \vasicek\ model is a one-factor model, it is well-known that the full (bond) opportunity set is covered by a single bond issue. For now, we will therefore consider the choice of $M_B$ to be arbitrary; the relation between investment horizon and bond maturity will be discussed further below.

It is shown in Appendix \ref{appendix_capmkt}, \eqref{capmkt_appendix_bond_dynamics}, that bond volatility, $\sigma_u^B$, is negative and given by 
\begin{equation}\label{bond_volatility}
	\sigma_u^B = -\psi_a(M_B-u) \sigma_r
\end{equation}
where $\psi_a(\cdot)$ is given by \eqref{psi}. Negativity follows from the trivial fact that bond prices decrease with increasing interest rates and vice versa, hence, the bond price-dynamics is given by
\begin{equation*}
	\frac{dB_u}{B_u} = (r + \xi_u^r \sigma_u^B) du + \sigma_u^B dW^r_u .
\end{equation*}

%% file: portfoliodyn.tex
\section{Portfolio Dynamics}\label{sec:portfolio_dynamics}
The risk-premium dynamics is an expression of the capital market opportunity set, hence, to derive the risk-return characteristics of a portfolio invested in the capital market, the next step is to define an investment strategy.

As discussed in the introduction, we consider investment strategies pre-committed to a specific asset allocation as time unfolds as practiced by pension providers and similar long-term investors, hence, investment strategies depend on time only. We therefore define an investment strategy, $\phi(u)$, starting at time $t$ with investment horizon $s>t$ as a continuous function, 
\begin{equation*}
	\phi(u) = (b(u),x_B(u),x_S(u))^T, 
\end{equation*}
on the interval $t \leq u \leq s$ where $x_B(u)$ is the number of bonds, $x_S(u)$, the number of shares at time $u$, and $b(u)$ is the holding in the bank account, $B_u$ at time $u$.

Following \cite{harrisonpliska1981}, the value, $V_u(\phi)$, of the investment strategy, $\phi$, at time $s > t$ subject to the self-financing conditions is then given as
\begin{equation}\label{portdyn_selffinancing}
	V_s(\phi) =  V_t + \int_t^s \big[ b(u) dB_u + x_B(u) dB_u + x_S(u) dS_u  \big]
\end{equation}
where $V_t>0$ is the initial portfolio value. The integral is well-defined since $\phi$ is trivially adapted and square integrable, hence, the investment strategy, $\phi$, is attainable and the portfolio value, $V_s(\phi)$, is well-defined.

Next, following \cite{bjork04} the value process is given by
\begin{equation*}
	V_u(\phi) = b(u) B_u + x_B(u) B_u + x_s(u) S_u
\end{equation*}
which - upon combination with the self-financing condition, \eqref{portdyn_selffinancing} - yields the following portfolio dynamics
\begin{align} \label{portdyn_general_dV}
	dV_u =  V_u r du + & b(u) dB_u + x_B(u) dB_u + x_S(u) dS_u \\
		& - \big[b(u) B_u + x_B(u) B_u + x_S(u) S_u) \big] r du \nonumber
\end{align}
where the shorthand $V_u(\cdot) \equiv V_u$ was introduced. 

At this stage, the model puts no restrictions on leverage nor shorting. In portfolio theory, it is customary to apply a non-negativity constraint to the allocation to each asset, one example would be the original model by Markowitz, \cite{markowitz52}, but it is well-known that such constraints are difficult to solve analytically and one must therefore often resort to numerical methods.

A less restrictive interpretation, though, is that the rationale for the no-shorting condition is to preclude portfolio loss beyond the initial capital. In this spirit we therefore rewrite \eqref{portdyn_general_dV} as
\begin{align} \nonumber
	\frac{dV_u}{V_u} & = r du + \frac{x_B(u)B_u}{V_u} \left(\frac{dB_t}{B_t} - r dt\right)
	+ \frac{x_S(u)S_u}{V_u} \left(\frac{dS_t}{S_t} - r dt\right) \\  \nonumber
	& = r du + \sum_{i = r,S} f_i(u)\Big(\xi^i(u)du + dW^i_u\Big)  , \quad V_u > 0, t \leq u \leq s  \\ \label{dV_t}
	& = (r_u + f_u^T \xi_u) du + f_u^T dW_u	
\end{align}
where the factor exposure, $f_u = (f_u^r,f_u^S)^T $, is given by
\begin{equation*}
	f(u) = V_u^{-1}\begin{pmatrix}
	x_B(u)B_u\sigma^B_u \\ x_S(u)S_u\sigma_S
	\end{pmatrix}
\end{equation*}
is the volatility of the allocation to bonds and equity, respectively. 

The following Proposition shows that these are sufficient conditions to ensure the investment risk never exceeds total loss of initial capital:

\begin{prop} \label{prop_nominal_meanvar} {\bf (Portfolio Dynamics)}
Given the assumptions of Proposition \ref{prop_xi_definition}, given initial values, $r_t, x_t$, at time $t$, and given the investment horizon $s>t$ and further assuming that the factor exposure, $f_u = (f^r_u, f^S_u)^T$, for $t \leq u \leq s$ is deterministic (depends on time only), then given the initial portfolio value, $V_t$, the horizon portfolio value, $V_s$, is given by
\begin{equation} \label{Vs}
	V_s = V_t \exp \left\{  \int_t^s  (r_u +  f_u^T \xi_u) du - \frac{1}{2} \int_t^s f_u^T \boldsymbol{C} f_u du + \int_t^s f_u^T dW_u \right\}
\end{equation}
 and is log-Normally distributed with conditional mean
\begin{equation} \label{nominal_mean}
	\E \log (V_s/V_t) = 
		 \int_t^s  \left\{ \epsilon_0 +  
		\Big( \epsilon_1 + f_u \Big)^T  \left(\xibar +  e^{-\boldGamma (u-t)} \xi_t \right) - \frac{1}{2} f_u^T \boldsymbol{C}  f_u 
		\right\} du 
\end{equation}
and conditional variance
\begin{equation} \label{nominal_variance}
	\V \log(V_s / V_t) = \int_t^s h_u^T \boldsymbol{C} h_u du
\end{equation}
where
\begin{equation} \label{h_nominal_definition}
	h_u = f_u + \boldxi \int_u^s  e^{-\boldGamma (v-u)}  (\epsilon_1 +  f_v)  dv
\end{equation}
and 
\begin{equation*}
	\boldC = \begin{bmatrix}
		1 & \rho \\ \rho & 1
	\end{bmatrix}, \qquad
	\epsilon_0 = \frac{ab - \rbar\kappa}{a - \kappa}, \qquad
	\epsilon_1 = \left(\frac{\sigma_r}{a-\kappa} , 0 \right)^T.
\end{equation*}

\begin{proof}
See appendix \ref{proof_nominal_meanvar}.
\end{proof}
\end{prop}

%% file: optimnom.tex
\section{Portfolio Optimization}\label{portfolio_optimization}
Given a factor allocation (investment strategy), $f_u$, Proposition \ref{prop_nominal_meanvar} provides horizon mean, \eqref{nominal_mean}, and horizon variance, \eqref{nominal_variance}, for any given investment horizon, $s$, hence, we have the necessary tools to proceed to search for {\em the} factor allocation that maximizes horizon expected return for fixed horizon variance.

It is important to stress, that factor allocation is allowed to change over time, hence, the optimial factor allocation for horizon $s$ cannot be expected to be an aggregate of optimal factor allocations over sub-periods: Each investment horizon provides a different opportunity set, hence, the optimal factor allocation explicitly depends on the investment horizon.

Before we move to the derivation of the optimality condition, it is convenient first to define the integral transformation, $y_u$, of the factor allocation, $f_u$, as
\begin{equation} \label{ynom_definition}
	y_u = \int_u^s   e^{-\boldGamma (v-u)}   f_v dv \quad \Rightarrow \quad 	\dot{y}_u = \frac{\partial y}{\partial u} = -  f_u + \boldGamma y_u
\end{equation}
from which it follows that $f_u$ can be reconstructed as
\begin{equation*} \label{fnom_definition}
	f_u  =  \boldGamma y_u - \dot{y}_u.
\end{equation*}
With this definition, we can re-parameterize the horizon mean, \eqref{nominal_mean}, to define the integral
\begin{multline} \label{Idefinition}
	I = \E \log (V_s/V_t) =  \int_t^s p(y_u, \dot{y}_u, u) du \\
	= \int_t^s  \Big\{ \epsilon_0 + 
	\Big( \epsilon_1 + \boldGamma y_u - \dot{y}_u \Big)^T  \left(\xibar +  e^{-\boldGamma (u-t)}\xi_t \right) 
	\\ - \frac{1}{2}  (\boldGamma y_u - \dot{y}_u)^T 	\boldsymbol{C} (\boldGamma y_u - \dot{y}_u) 
	\Big\} du 
\end{multline}
and similarly, we can re-parameterize horizon variance, \eqref{nominal_variance}, by combining \eqref{h_nominal_definition} and \eqref{ynom_definition} to define the integral
\begin{equation*} \label{Jdefinition}
	J = \V \log(V_s / V_t)  = \int_t^s q(y_u, \dot{y}_u, u) du = \int_t^s h_u^T \boldsymbol{C} h_u du
\end{equation*}
where $h_u$ is now given by
\begin{align}   \label{thm_nominal_hu_definition}
	h_u & = f_u + \boldxi \int_u^s  e^{-\boldGamma (v-u)}  (\epsilon_1 +  f_v)  dv \\  \nonumber
	& = \underbrace{\boldGamma y_u - \dot{y}_u}_{f_u} + \int_u^s  e^{-\boldGamma (v-u)} \boldxi \epsilon_1  dv
	+ \boldxi \underbrace{\int_u^s  e^{-\boldGamma (v-u)} f_v  dv }_{y_u} \\ \nonumber
	& = \boldGamma_\xi y_u - \dot{y}_u + \psi_{\kappa}(s-u) \eta^r
\end{align}
with $\boldGamma_\xi = \boldGamma + \boldxi$, and where $\eta^r = (\sigma_r, 0)^T$.

\subsection{The Euler-Lagrange\ Equation}\label{sec:euler_lagrange}
Given this new parameterization, mean-variance optimization is equivalent to maximize $I$ for fixed $J$ which - following the original idea of \cite{jarner21} - can be solved by classical methods of Calculus of Variations.

By introducing the Lagrange\ multiplier, $\nu$, the mean-variance optimization criteria therefore is recast into optimizing
\begin{equation*}
	I^* =  I - \nu J /2
\end{equation*}
where - in the tradition of utility optimization - the optimization problem is stated as the disutility in $\nu$, hence, we will from here refer to $\nu$ as the {\em risk aversion parameter} since it $I^*$ balances the objective of maximizing the expected return, $I$, at the 'cost' of variance, $J$. For this ratio to be positive, risk aversion therefore must be positive corresponding to the upper branch of the efficient frontier in traditional mean-variance optimization, whereas for $\nu<0$, the marginal gain is negative corresponding to the lower branch of the efficient frontier.

Furthermore, the limit $\nu \rightarrow \infty$ is the global minimum for horizon mean return whereas $\nu \rightarrow 0+$ is the free optimization problem subject to the maximum-loss restriction discussed in Section \ref{sec:portfolio_dynamics}.

Following \cite{weinstock}, $I^*$ is optimized by the (re-parameterized) factor allocations $y_u$ satisfying the Euler-Lagrange\ equation
\begin{equation} \label{thm-nominal-general-Euler-lagrange}
	\frac{\partial p_u^*}{\partial y_u} - \frac{d}{du} \left( \frac{\partial p_u^*}{ \partial \dot{y}_u} \right) = 0
\end{equation}
where  $p^*(y_u, \dot{y}_u, u) = p(y_u, \dot{y}_u, u) - \nu q(y_u, \dot{y}_u, u)/2$ and $p(\cdot)$ is defined in \eqref{Idefinition} and $q(\cdot)$ is defined in \eqref{Jdefinition}.

Care must be taken in setting the boundary conditions: At the initial time, $u=t$, the optimization is free, hence, the lower boundary is a transversal - or natural - boundary condition problem whereas the upper limit is dictated by the definition of $y_u$, \eqref{ynom_definition}, which by construction must be zero at the upper boundary, i.e., $y_s = 0$. In summary, the boundary conditions are
\begin{subequations}
\begin{align} \label{thm-nominal_lower_boundary_condition}
	\frac{\partial p_u^*}{ \partial \dot{y}_u}\Big|_{u=t} &  = 0 \qquad & \text{(lower boundary)} \\ \label{thm-nominal_upper_boundary_condition}
	y_u|_{u=s} & = 0 & \text{(upper boundary)}
\end{align}
\end{subequations}
for $u=t$ and $u=s$. We summarize the result in the following Theorem:
\begin{thm}\label{theorem_nominal_Euler}
{\bf (Mean-Variance Optimization)}
Given the assumptions and definitions of Propositions \ref{prop_xi_definition} and \ref{prop_nominal_meanvar} then the (deterministic) investment strategy, $f_u = (f^r_u, f^S_u)^T$, maximing the conditional expected nominal return, $\mathbb{E} \log (V_s/V_t)$, for fixed (constant) conditional horizon variance, $\mathbb{V} \log (V_s/V_t)$, is given by
\begin{equation*} \label{nominal_euler_fu_definition}
	f_u = \boldGamma y_u - \dot{y}_u, \quad \mbox{for $t \leq u \leq s$}
\end{equation*}
where $y_u$ is the solution to the inhomogeneous second-order (matrix) differential equation
\begin{equation} \label{thm_nominal_equation}
	(1+\nu) \Big[ \boldC\ddot{y}_u + \boldB \dot{y}_u - \boldA y_u \Big] = g_u
\end{equation}
and
\begin{equation*}
	\boldA  = \begin{Bmatrix}
		\gamma_r^2 & 0 \\ 
		0 & \gamma_S^2
	\end{Bmatrix}
	+ \rho \begin{Bmatrix}
		0 & a_\nu  \\
		a_\nu & 0
	\end{Bmatrix} \qquad
	\boldB  = \rho \begin{Bmatrix}
	0 & b_{\nu} \\
	-b_{\nu} & 0
\end{Bmatrix} 
\end{equation*}
with
\begin{align} \nonumber
	&\gamma_r^2  =  \frac{\kappa^2  +\nu  a^2 }{1 +\nu } && \gamma_S^2  = \frac{\alpha^2  +\nu  (\alpha')^2}{1 +\nu } \\ \nonumber
	&a_\nu  =  \frac{\alpha\kappa +\nu a \alpha'}{1 +\nu } && b_\nu  =\frac{(\kappa-\alpha)  +\nu (a-\alpha')}{1 +\nu }
\end{align}
where $\alpha' = \alpha - \sigma_x/\sigma_S$ and
\begin{equation*}
	g_u  =  -\Big[ 
	\begin{pmatrix}
		\kappa\xi_0^r \\ \alpha \xi_0^S
	\end{pmatrix}
	 +\nu \sigma_r
	\begin{pmatrix}		
		1 - (\kappa+a) \psi_\kappa(s-u) \\ \rho [1-(\kappa+\alpha')\psi_\kappa(s-u)]
	\end{pmatrix}\Big]
\end{equation*}
and where $0<\nu<\infty $ is the risk aversion coefficient.\newline
Furthermore, $y_u$ satisfies the boundary conditions
\begin{align*}% \label{nominal_lower_boundary_condition}
	& b_0 + \big( b_u - [ \boldGamma  +\nu  \boldGamma_\xi ] y_u + (1 +\nu )\dot{y}_u \big) \big|_{u = t} = 0 & \text{\rm (lower boundary)}\\ %\label{nominal_upper_boundary_condition}
	& y_u |_{u=s} = 0 & \text{\rm (upper boundary)}
\end{align*}
at times $t$ and $s$, respectively, where
\begin{equation*}
	\boldGamma_\xi = \boldGamma + \boldxi, \qquad
	\boldC b_0 = \xibar, \qquad
	\boldC b_u = e^{-\boldGamma(u - t)} \xi_t 
	-\nu\sigma_r \psi_\kappa(s-u)
	\begin{pmatrix}
		1 \\ \rho
	\end{pmatrix}.
\end{equation*}

\begin{proof}
See appendix \ref{proof_theorem_nominal_Euler}.
\end{proof}
\end{thm}
Explit results for the horizon mean and horizon variance are provided in Lemma \ref{lem:meanvar} in the Appendix.

%% file: spectral_problem.tex
\section{The Spectral Problem} \label{spectral_solution}
Theorem \ref{theorem_nominal_Euler} shows that the (transformed) optimal factor allocation is the solution to an inhomogeneous second-order differential equation, hence, the solution - following standard arguments - is the sum of a particular and the homogeneous solution to \eqref{thm_nominal_equation}. In our case, we will consider homogeneous solutions of the $y^h_u = e^{\boldS(s-u)} k$, where $\boldS \in \R^{2\times 2}$, $k \in \R^2$, hence, the characteristic equation becomes a matrix equation
\begin{equation} \label{matrix_polynomial}
	P(\boldS) = \boldC \boldS^2 - \boldB \boldS - \boldA = 0
\end{equation}
(notice the change of sign of $\boldB$) where $P$ is a second order matrix polynomial with coefficients given by Theorem \ref{theorem_nominal_Euler}. Alas, the characteristic equation cannot be solved by standard (scalar) methods. Be also aware that generally, there will be more than two solutions to \eqref{matrix_polynomial} but, as will be shown below, under mild additional assumptions, unique pairs of solutions exist which will suffice for our purposes. Following \cite{dennis73} we will refer to solutions of \eqref{matrix_polynomial} as {\em solvents} of $P$.

In order to determine a suitable pair of solvents, we follow \cite{dennis73} and first define the lambda-matrix, $\boldM_\lambda$, as
\begin{equation}\label{lambda_matrix}
	\boldM_\lambda = P(\lambda \unitI) = \boldC \lambda^2 - \boldB \lambda - \boldA
\end{equation}
where $\lambda \in \C$ is a scalar in order to determine the {\em latent} roots of $P$, that is, values of $\lambda$ of which the lambda matrix, $\boldM_\lambda$, is degenerate:
\begin{equation} \label{lambda_matrix_determinant}
	\det \boldM_\lambda = \det \left( \boldC \lambda^2 - \boldB \lambda - \boldA\right) = 0.
\end{equation}
Since $\boldA, \boldB, \boldC \in \R^{2\times 2}$ this amounts to determining the roots of a 4'th-order scalar polynomial. 

The problem of determining the latent roots can be simplified, though: By the properties of the coefficients of $\boldM$, cf.\ Theorem \ref{theorem_nominal_Euler}, and by the properties of the determinant, it follows that
\begin{equation*}
	\det \boldM_\lambda = \det \boldM_\lambda^T = \det \left( \boldC (-\lambda)^2 - \boldB (-\lambda) - \boldA\right)  = 0,
\end{equation*}
hence, if $\lambda$ is a latent root, then $-\lambda$ is also a latent root, that is, \eqref{lambda_matrix_determinant} is a quadratic (scalar) polynomial in $\lambda^2$. The full set of latent roots can therefore be written as $\lambda = (\lambda_1, \lambda_2, -\lambda_1, -\lambda_2)$ where we for now will assume the four roots to be distinct (the exact condition is given in Theorem \ref{thm_spectral_problem} below).

Since by definition, $\boldM_\lambda$ is singular at the latent roots, the equation
\begin{equation} \label{latent_right_vector}
	\boldM_{\lambda_i} r_i = 0
\end{equation}
has a non-trivial solution for each latent root. The set $\{ r_1,\ldots, r_4 \}$ is called the {\em right latent vectors} of $\boldM_\lambda$ and cannot - in general - be assumed to be distinct but by Theorem 4.1 of \cite{dennis73} every linearly independent pair of right latent vectors, $\{r_i, r_j\}$, provides a (right) solvent,
\begin{equation} \label{right_solvent}
	\boldS_{ij} = \boldQ_{ij} \boldLambda_{ij} \boldQ_{ij}^{-1},
\end{equation}
of $P$ where $\boldLambda_{ij} = {\rm diag}(\lambda_i,\lambda_j)$ and $\boldQ = \{r_i; r_j \}$ is a square matrix of full rank with columns $r_i$ and $r_j$, respectively. A set of solvents is called complete if the eigenvalues exactly coincide with the latent roots.

To see the solvent, \eqref{right_solvent}, is a solution, insert \eqref{right_solvent} into \eqref{matrix_polynomial}:
\begin{align*}
	P(\boldS_{ij}) & = \boldC \boldQ_{ij} \boldLambda_{ij}^2 \boldQ_{ij}^{-1}  - \boldB \boldQ_{ij} \boldLambda_{ij} \boldQ_{ij}^{-1} - \boldA \\
	& = \Big[
	\boldC \{ \lambda_i^2 r_i; \lambda_j^2 r_j \}
	- \boldB \{ \lambda_i r_i; \lambda_j r_j \} - \boldA \{ r_i; r_j \}  \big] \boldQ_{ij}^{-1} \\
	& = \Big\{ \big[ \boldC \lambda_i^2 - \boldB \lambda_i - \boldA  \big] r_i; \big[ \boldC \lambda_j^2 - \boldB \lambda_j - \boldA  \big] r_j \Big\}Q_{ij}^{-1}
\end{align*}
which column by column is zero by \eqref{latent_right_vector} .

Corollary \ref{cor:corollary_latent_roots} provides the condition for latent roots to be distinct as well as provides explicit formulas for latent roots, latent vectors, and solvents. We summarize our findings in the following theorem:
\begin{thm} \label{thm_spectral_problem}
	{\bf (Spectral Problem)} Given the assumptions of Theorem \ref{thm_spectral_problem}  and if the discriminant
	\begin{equation*} \label{discriminant_latent_roots}
		D = (1-\rho^2)\big[\gamma_r^2 - \gamma_S^2\big]^2 + \rho^2 \big[ \gamma_r^2 + \gamma_S^2 - (2a_{\nu}+b_{\nu}^2) \big]^2 - \rho^2 (1-\rho^2) (4a_{\nu} 	+b_{\nu}^2)b_{\nu}^2
	\end{equation*}
	is non-zero, then there exists a pair of latent roots, $(\lambda_1, \lambda_2) \in \C^2$, such that the set of solvents, $\boldS_1, \boldS_2$, of \eqref{matrix_polynomial} is real and complete and is given by
	\begin{equation*}
		\boldS_1 = \boldQ_1 \boldLambda \boldQ_1^{-1}, \qquad \boldS_2 = -\boldQ_2 \boldLambda \boldQ_2^{-1}, \qquad \boldS_1,\boldS_2\in \R^{2\times 2}
	\end{equation*}
	where $\boldLambda = {\rm diag}(\lambda_1, \lambda_2)\in \C^{2\times 2}$ and $\boldQ_1, \boldQ_2 \in \C^{2\times 2}$, are square matrices where columns are right-latent vectors of $(\lambda_1,\lambda_2)$ and $(-\lambda_1,-\lambda_2)$, respectively.
	\begin{proof}
		This follows directly Theorems 4.1 and 4.5 in \cite{dennis73} and Corollary \ref{cor:corollary_latent_roots}.
	\end{proof}
\end{thm}

%% file: general_solution.tex
\section{General Solution}\label{sec:solution}
By standard arguments, the transformed optimal factor allocation, $y_u$, is the sum,
\begin{equation*}
	y_u = y^p_u + y^h_u,
\end{equation*}
of a particular, $y^p_u$, and the homogeneous, $y^h_u$, solutions to \eqref{thm_nominal_equation}, respectively. Theorem \ref{thm_spectral_problem} provides the homogeneous solution and the particular solution and implications of the boundary conditions, \ref{thm-nominal_lower_boundary_condition} and \ref{thm-nominal_upper_boundary_condition}, are given in Appendix \ref{proof_nominal_solution}. The general solution is stated in the following Theorem:
\begin{thm} \label{theorem_nominal_solution}
	{\bf (General Solution)} Given the assumptions of Theorem \ref{theorem_nominal_Euler} and Theorem \ref{thm_spectral_problem} then the optimal factor allocation, $f_u$, is given by
	\begin{equation}\label{thm:fu}
		f_u = (\boldGamma k_1 + k_2) + (\boldGamma + \boldS_1) e^{\boldS_1 (s-u)} q_{1t} + (\boldGamma + \boldS_2) e^{\boldS_2 (s-u)} q_{2t}, \quad t \leq u \leq s
	\end{equation}
	where $k_1, k_2 \in \R^2$ are given by
	\begin{equation*}
		k_2 = \Big(
		\frac{\sigma_r}{\kappa-a} , 0
		\Big)^T
		\qquad
		(1 +\nu )\boldA k_1  = \begin{pmatrix}
			\kappa \xibar^r \\ \alpha \xibar^S
		\end{pmatrix}
		+ \frac{\sigma_r}{a-\kappa}
		\begin{pmatrix}
			\kappa  +\nu  a \\ \rho (\alpha  +\nu  \alpha')
		\end{pmatrix}
	\end{equation*}
	and $q_{1t}, q_{2t} \in \R^2$ as the solution to
	\begin{equation*}
		\begin{Bmatrix}
			\unitI & \unitI
			\\
			\boldD_1  e^{\boldS_1 (s-t)} & \boldD_2  e^{\boldS_2 (s-t)}
		\end{Bmatrix} 
		\begin{pmatrix}
			q_{1t} \\ q_{2t}
		\end{pmatrix} 
		= \begin{pmatrix}
			-k_1 \\
			\boldC^{-1} (\xibar + \xi_t) - (\boldGamma k_1 + k_2)  -\nu (\boldGamma_\xi k_1 + k_2)
		\end{pmatrix}.
	\end{equation*}
	where $\boldD_1 = \boldGamma  +\nu \boldGamma_\xi + (1 +\nu )\boldS_1$ and $\boldD_2 = \boldGamma  +\nu \boldGamma_\xi + (1 +\nu )\boldS_2$.
	
	\begin{proof}
		See Appendix \ref{proof_nominal_solution}.
	\end{proof}	
\end{thm}

We see that the optimal strategy is a linear combination of exponentials of the two solvents, $\boldS_1$ and $\boldS_2$. To gain a little more insight into the structure of the solution, from Theorem \ref{thm_spectral_problem} we write for an arbitrary term
\begin{equation*}
	e^{\boldS(s-u)} = \boldQ \begin{bmatrix}
		e^{\lambda_i(s-u)} & 0 \\ 0 & e^{\lambda_j(s-u)}
	\end{bmatrix}
	\boldQ^{-1},
\end{equation*}
that is, the optimal factor allocation explicitly depends on the investment horizon when the inverse of (at least one of) the latent roots, $|\lambda_{\cdot}|^{-1}$, is of the order of the investment horizon.

Elaborating on this, consider two investors starting at different times $t_2>t_1$, respectively, but with the same investment horizon date, $s$: The first investor would commit to a deterministic investment strategy at $t_1$ from prevailing market conditions at that time, that is, the state of the market as expressed by $\xi_{t_1}$. When the second investor enters the market at $t_2$, market conditions would have changed and the investor would commit to a different investment strategy than the first investor even if they agree on model and parameters, hence, are solving the same spectral problem.

Finally, a particular property of the model is, that the optimization is over a complete bond market. It is therefore an endogeneous outcome that in the limit of infinite risk aversion, $\nu \rightarrow \infty$, the optimal factor strategy is to buy (and hold) the zero-coupon bond maturing at the investment horizon to limit horizon variance to zero as stated in the following Corollary:
\begin{cor}\label{cor:infinite_riskaversion}
	{\bf (Infinite Risk-Aversion)} Given the assumptions of Theorem \ref{theorem_nominal_solution} then the optimal factor allocation, $f^\infty_u$, in the limit of infinite risk-aversion, $\nu \rightarrow \infty$, is given by
	\begin{equation*}
		f^\infty_u = \begin{pmatrix}
			-\psi_a(s-u)\sigma_r \\ 0
		\end{pmatrix}
	\end{equation*}
	 and coincides with buy-and-holding a zero-coupon bond maturing at $s$ where $\psi_a(\cdot)$ is given by \eqref{psi}.
	
	\begin{proof}
		See Appendix \ref{proof:infinite_riskaversion}.
	\end{proof}	
\end{cor}

%% file: discussion.tex
\section{Discussion}\label{sec:discussion}
\begin{figure}[b]
	\caption{Efficient Frontier by Investment Horizon}\vspace{0.5cm}
	\includegraphics[width=1\linewidth]{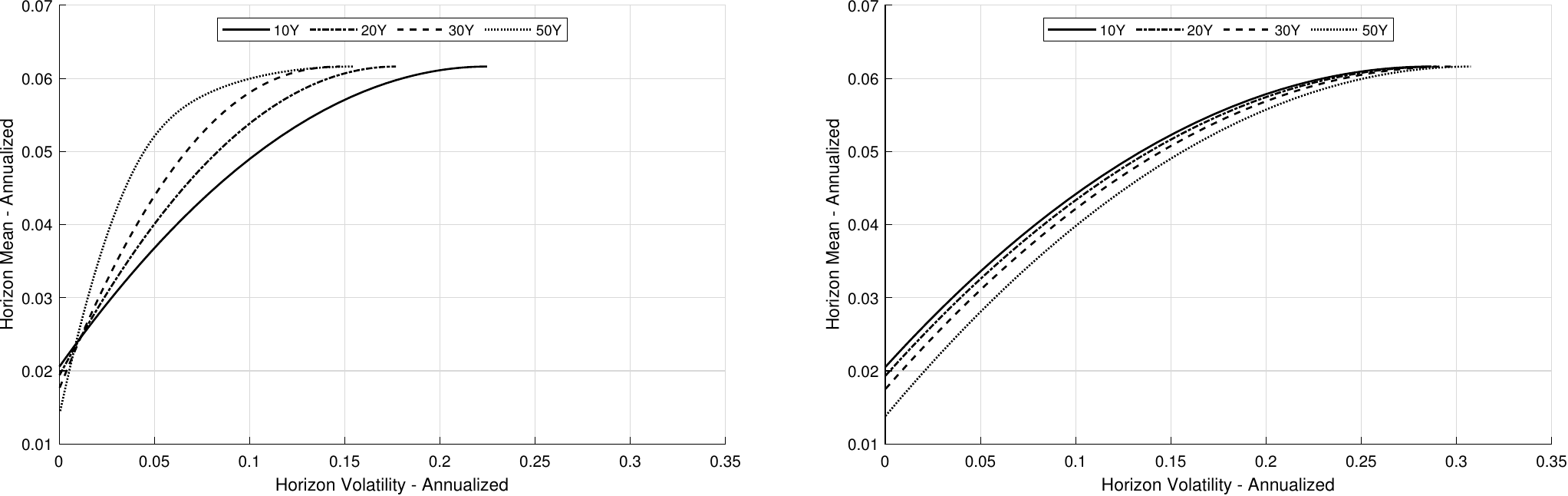}
	\raggedright \small \em Annualized horizon volatility vs annualized horizon mean return for investment horizons 10Y, 20Y, 30Y, and 50Y. Each efficient frontier is parametrized by risk aversion, $\nu$, and is plottet for the interval: $\nu \in (0, \infty)$. Illustrative parameters are given in Table \ref{table:parameters}. (Left) Slow mean-reversion in equity risk-premium ($\alpha=0.01$). (Right) Fast mean-reversion in equity risk-premium($\alpha=0.25$); all other parameters unchanged.
	\label{fig:efficient_frontier}
\end{figure}
Efficient frontiers are illustrated in Figure \ref{fig:efficient_frontier} for two illustrative sets of capital market assumptions, cf.\ Table \ref{table:parameters}: The first set of parameters, (left), assumes slow mean-reversion in the equity risk-premium, hence, holding equity for longer periods of time is rewarded as mean-reversion manifests itself. The second set, (right), assumes fast (no) mean-reversion in the equity risk-premium, hence, there is no advantage in holding equity for longer periods of time to short periods.

In either case, efficient frontiers start in the same point for zero volatility, since zero horizon volatility can only be realized by allocation to a zero-coupon bond maturing at the investment horizon. The variation across investment horizon reflects the bond risk-premium as given by the (\vasicek) term structure is different by maturity whereas the initial point is independent of the equity risk-premium.

For less risk-averse investors, the risk-return trade-off is markedly different if the equity risk-premium is mean-reverting slowly or not: In the latter case, (fast mean-reversion) short-term and long-term investors are essentially offered the same risk-return trade-off, hence, efficient frontiers converge to the same point.

If the equity risk-premium mean-reverts slowly, long-term investors are rewarded by the ability to hold risky positions with a much better risk-return trade-off than offered shorter-term investors: The slope of the efficient frontier is particularly steep for small levels of risk and it is clear that long-term investors would benefit materially by moving from no risk to just a little risk in the portfolio. This potentially has huge implications to pension funds and other asset managers obliged to provide stable income streams because assuming just a little risk materially brings down the cost of such products.

\subsection{Factor Allocation}
\begin{figure}[b]
	\caption{Factor Allocation by Risk-Aversion}\vspace{0.5cm}
	\includegraphics[width=1\linewidth]{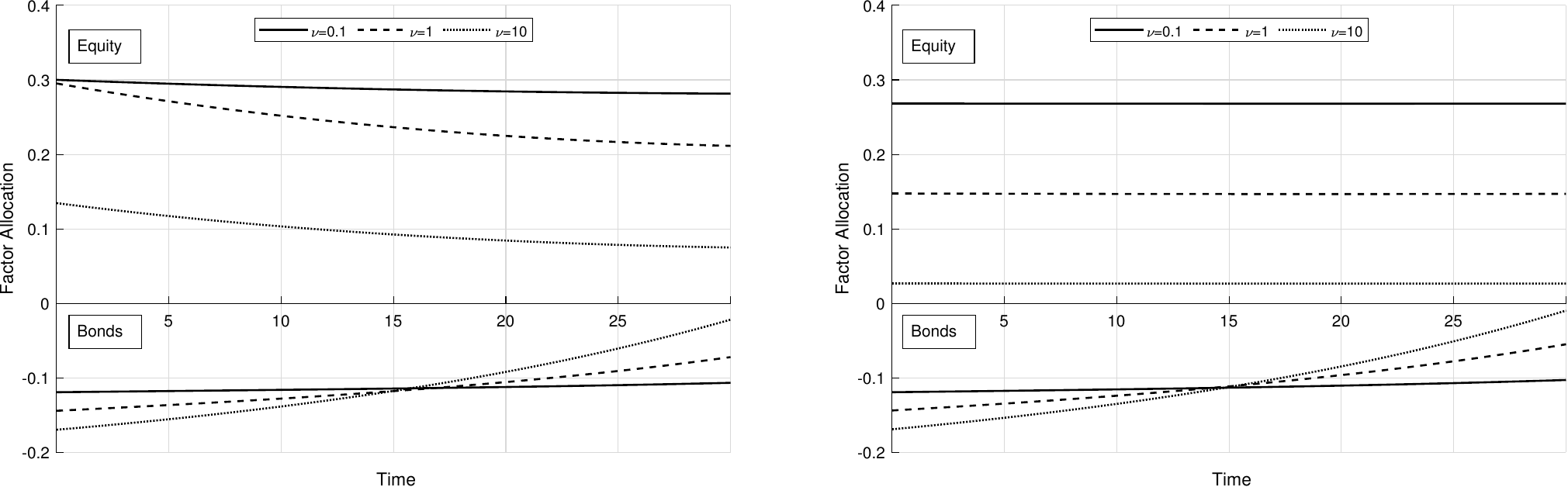}
	\raggedright \small \em Factor allocation to equity (positive) and bonds (negative) at 30Y investment horizon for horizon-volatility set by risk averson $\nu=0.1,1,10$. The factor allocation provides the deterministic target volatility of each asset class over time which by \eqref{bond_volatility} is negative for bonds. (Left) Factor allocation assuming slow mean-reversion and (right) assuming fast (no) mean-reversion in the equity risk-premium. Parameters are given in Table \ref{table:parameters}.
	\label{fig:factor_allocation}
\end{figure}

The deterministic factor allocation over the investment horizon is illustrated in Figure \ref{fig:factor_allocation} for three levels of risk aversion, $\nu = 0.1,1,10$, assuming slow (left) and fast (right) mean-reversion in the equity risk-premium.

Starting with the case of fast mean-reversion (right) in the equity risk-premium, we see that allocation to equity for all practical purposes is constant over the investment horizon in support of well known practices of re-balancing portfolio risk over time. For bonds, this is only true for low levels of risk-aversion when terminal volatility is high. For higher levels of risk-aversion, the allocation to bond is decreasing in absolute terms with the investment horizon. Bond volatility is proportional to bond duration, hence, the declining volatility is a reflection of some horizon variance being hedged by the matching bond.

For slow mean reversion (left) in the equity risk-premium, first notice that generally equity levels are higher than in the 'fast' case: This is because the negative serial correlation introduced by the mean-reverting equity risk-premium results in a narrower distribution in equity returns, hence, a higher level of equity can be sustained relative to the alternative case. Moreover, the equity allocation tend to be declining over time when the equity risk-premium mean-reverts slowly. 

This is in contrast to bond allocations, which more or less are identical to the alternative case. Also in this case is bond volatility declining over time as horizon volatility is hedged more and more by bonds with increasing risk aversion.

%% file: conclusion.tex
\section{Conclusion}\label{sec:conclusion}
Being a long-term investor has become an argument by itself for holding larger allocations to risky assets despite little theoretical support for this argument. In many applications, it is argued that investor preferences depend on investor age, hence, arguments more have the character of being postulate than rationale as to why capital market would provide a different opportunity set to investors with different investment horizons.

In contrast, the objective of this paper is to formulate a theoretical foundation in which such investment beliefs are reflected directly in the capital market opportunity set and to derive proper mean-variance efficient asset allocations from these assumptions. It was shown, that long-term investors indeed can justify higher allocations to risky assets if risk-premia are slowly mean-reverting whereas if risk-premia are not, there is no particular argument that being a long-term investor poses a different allocation problem than to any other investor.

%% file: capmkt_appendix.tex
\section{Capital Market Model}\label{appendix_capmkt}

\begin{table}[b]
	\caption{Illustrative Capital Market Parameters}
	\centering
	\begin{tabular}{l|c|c|c}
		Description & Parameter & \multicolumn{2}{c}{Mean-Reversion} \\
		& & Slow & Fast (None) \\
		%		\hhline{=|=|=|=}
		\hline
		Interest Rate - P-measure & &  &  \\
		\hline
		Mean-Revertion Strength & $\kappa$ & 0.05 & 0.05 \\
		Mean Level & $\rbar$ & 0.02 & 0.02 \\
		Volatility & $\sigma_r$ & 0.01 & 0.01\\
		\hline
		Interest Rate - Q-measure & &  &  \\
		\hline
		Mean-Reversion Strength & $a$ & 0.04 & 0.04 \\
		Mean Level & $b$ & 0.03 & 0.03 \\
		\hline
		Equity Risk-Premium & &  &  \\
		\hline
		Mean-Revertion Strength & $\alpha$ & 0.01 & 0.25\\
		Mean Level & $\xbar$ & 0.04 & 0.04\\
		Volatility & $\sigma_x$ & 0.007  & 0.007\\
		\hline
		Equity & &  &  \\
		\hline
		Volatility & $\sigma_S$ & 0.15 & 0.15\\
		Correlation With Interest Rate & $\rho$ & 0.25 & 0.25 \\
	\end{tabular}
	\raggedright \small \em \vspace{0.5cm} \newline Illustrative capital market parameters for slow ($\alpha=0.01$) and fast ($\alpha=0.25$) mean-reversion in the equity risk-premium. Notice that interest-rate parameters are chosen to ensure a negative risk-premium for bonds, cf.\ \eqref{bond_volatility}.
	\label{table:parameters}
\end{table}

Given time-$t$ values of the variables: $(r_t, x_t, S_t)$, then for $s>t$, we quote select integral representations of \eqref{dS_t} to \eqref{dr_t} without proof from \cite{jp17}
\begin{subequations}
\begin{align} 
	r_s & = \bar{r} + e^{-\kappa (s-t)}(r_t - \bar{r}) + \sigma_r \int_t^s e^{-\kappa (s-u)} dW^r_u, \label{r_s}, \\
	x_s & = \bar{x} + e^{-\alpha (s-t)}(x_t - \bar{x}) - \sigma_x \int_t^s e^{-\alpha(s-u)} dW^S_u, \label{x_s}
\end{align}
\end{subequations}
with mean 
\begin{subequations}
\begin{align}
	\E r_s  &= \rbar + e^{-\kappa (s-t)} (r_t-\rbar) \\ 
	\label{capmkt_cash_mean}
	\E \int_t^s r_u du  &= \rbar(s-t) + \psi_\kappa(s-t)(r_t - \rbar) 
\end{align}
\end{subequations}
and variance
\begin{subequations}
	\begin{align}
	\V r_s &= \sigma_r^2 \psi_{2\kappa}(s-t) \\ 
	\label{capmkt_cash_variance}
	\V \int_t^s r_u du &= \sigma_r^2 \upsilon_\kappa(s-t) 
\end{align}
\end{subequations}
where
\begin{align} \label{capmkt_psi}
	\psi_\alpha(s-t) & = \int_t^s e^{-\alpha (u-t)} du = \frac{1}{\alpha} \big(1-e^{-\alpha(s-t)} \big) \\ \nonumber
	\upsilon_\alpha(s-t) & = \int_t^s \psi_\alpha^2(u-t)du = \frac{(s-t) - 2\psi_{\alpha}(s-t) + \psi_{2\alpha}(s-t) }{\alpha^2} .
\end{align}
The full covariance matrix is given in \cite{jp17}.

\subsection{Nominal Bonds}
Following \cite{jp17}, the price, $p_t(s)$, at time $t$ of a (nominal) zero-coupon bond maturing at time $s$ is given by the Vasicek formula
\begin{align*}
	p_t(s) & = e^{-R_t(s) (s-t)}
	\intertext{where the zero-coupon rate, $R_t(s)$, is given by} \label{capmkt_appendix_zcrate}
	R_t(s) & = b + \frac{r_t - b}{s-t} \psi_a(s-t) - \frac{\sigma_r^2}{2(s-t)}\upsilon(a,s-t)
	\intertext{and} 
	\upsilon(a,\tau) & =  \frac{\tau -2\psi_a(\tau) + \psi_{2a}(\tau)}{a^2}.
\end{align*}
By Itô's Lemma the temporal dynamic is
\begin{equation} \label{capmkt_appendix_bond_dynamics} 
	dp_t(s) = p_t(s) \big\{[r_t-\lambda_t^r \psi_a(s-t)\sigma_r] dt - \psi_a(s-t)\sigma_r dW^r_t \big\},
\end{equation}
hence, a positive excess return corresponds to a negative risk premium, $\lambda^r$.

%% file: prop_xidef.tex
\section{Proof of Proposition \ref{prop_xi_definition}} \label{proof_prop_xi_definition}
{\em Proof}. By direct insertion of \eqref{r_s} and \eqref{x_s} into \eqref{xi_t} it follows
\begin{equation}\label{prop_xi_def}
	\xi_s  =
	\begin{pmatrix}
		\left[(a-\kappa)\{ \bar{r} + e^{-\kappa (s-t)}(r_t - \bar{r}) + \sigma_r \int_t^s e^{-\kappa (s-u)} dW^r_u \}+ \kappa \rbar - ab\right]/\sigma_r \\
		\left\{\bar{x} + e^{-\alpha (s-t)}(x_t - \bar{x}) - \sigma_x \int_t^s e^{-\alpha(s-u)} dW^S_u \right\} /\sigma_S 
	\end{pmatrix}
\end{equation}
which by defining
\begin{equation*}
	\boldGamma  = \begin{Bmatrix} \kappa & 0  \\ 0 & \alpha \end{Bmatrix}
\end{equation*}
is rewritten into
\begin{multline*}
	\xi_s  = \underbrace{
	\begin{pmatrix}
		a(\bar{r} - b)/\sigma_r \\
		\bar{x}/\sigma_S 
	\end{pmatrix} }_{\xibar}
	+
	e^{-\boldGamma (s-t)} \underbrace{
	\begin{pmatrix}
		(a-\kappa)(r_t - \bar{r})/\sigma_r \\
		(x_t - \bar{x})/\sigma_S
	\end{pmatrix} }_{\xi_t}\\
	+ \underbrace{
	\begin{Bmatrix}
		(a-\kappa) & 0\\
		0 & - \sigma_x/\sigma_S 
	\end{Bmatrix} }_{\boldxi}	\int_t^s e^{-\boldGamma (s-u)} dW_u
\end{multline*}
from which Proposition \ref{prop_xi_definition} follows. 

Since the integrand of the stochastic integral is deterministic, it follows that $\xi_s$ is Normally distributed with conditional mean
\begin{equation*}
	\E \xi_{s|t} = \xibar + e^{-\boldGamma (s-t)} \xi_t
\end{equation*}
and conditional variance
\begin{equation*}
	\V \xi_{s|t} = \boldxi \int_t^s e^{-\boldGamma (s-u)} \boldC e^{-\boldGamma (s-u)} du \, \boldxi = \boldxi \boldsymbol{V}^\xi_{s|t} \boldxi
\end{equation*}
where from the definition \eqref{capmkt_psi}
\begin{equation*}
	\boldsymbol{V}^\xi_{s|t} = \begin{Bmatrix}
		\Psi(2\kappa,s-t) & \rho\Psi(\kappa+\alpha,s-t) \\
		\rho\Psi(\kappa+\alpha,s-t) & \Psi(2\alpha,s-t) 
	\end{Bmatrix}.
\end{equation*}
For $s \rightarrow \infty$, we see that $\xibar$ indeed is the asymptotic mean whereas the asymptotic variance, $\V \xi_\infty$, is given by $\V \xi_\infty = \boldxi \boldsymbol{V}^\xi_\infty \boldxi$ with
\begin{equation} \label{asymptotic_variance_xi_s}
	\boldsymbol{V}^\xi_\infty = \begin{Bmatrix}
		1/(2\kappa) & \rho/(\kappa+\alpha) \\
		\rho/(\kappa+\alpha) & 1/(2\alpha) 
	\end{Bmatrix}.
\end{equation} $\square$

%% file: prop_nominal_meanvar.tex
\section{Proof of Proposition \ref{prop_nominal_meanvar}} \label{proof_nominal_meanvar}
{\em Proof}. Let $g(x) = e^x$ and for $s\geq t$ let
\begin{equation} \label{Xs_definition}
	X_s = \int_t^s  (r_u +  f_u^T \xi_u) du - \frac{1}{2} \int_t^s f_u^T \boldsymbol{C} f_u du + \int_t^s f_u^T dW_u
\end{equation}
then it is straightforward by the multi-dimensional Itô's Lemma, cf.\ Proposition 4.18 of \cite{bjork04}, to show that the derivative of $V_s = V_t \, g(X_s)$ is given by
\begin{equation} \label{prop_meanvar_dVs}
	dV_s  = V_s \Big\{ (r_u +  f_u^T \xi_u) dt + f_u^T dW_s \Big\} .
\end{equation}
Since $V_s = V_t g(X_s)$ and since \eqref{prop_meanvar_dVs} is identical to \eqref{dV_t}, it follows that $V_t g(X_s)$ is the solution to \eqref{dV_t}, hence
\begin{equation*}
	V_s = V_t \exp \left\{  \int_t^s  (r_u +  f_u^T \xi_u) du - \frac{1}{2} \int_t^s f_u^T \boldsymbol{C} f_u du + \int_t^s f_u^T dW_u \right\}
\end{equation*}
as stated. $\square$

\subsection{Mean and Variance}
From the definition of $X_s$, \eqref{Xs_definition}, it follows from \eqref{r_s}, \eqref{prop_xi_def}, and \eqref{Xs_definition}, that the first term of $X_s$ is (a sum of) Itô-integrals of the structure
\begin{equation*}
	I = \int_t^s g(u) du \int_t^u e^{-\alpha(u-v)} dW_v
\end{equation*}
where $g(u)$ is a continuous, deterministic function and $W_v$ a Brownian motion. By interchanging the order of integration, we find
\begin{equation*}
	I = \int_t^s dW_v \int_v^s g(u) e^{-\alpha(u-v)} du = \int_t^s G(v) dW_v
\end{equation*}
where $G(v)$ also is a continuous, deterministic function, hence, the first term has the same structure as the last term. 

Furthermore, from Proposition \ref{prop_xi_definition} and upon defining 
\begin{equation*}
	\epsilon_0 = \frac{ab - \rbar\kappa}{a - \kappa}, \qquad
	\epsilon_1 = \left(\frac{\sigma_r}{a-\kappa} , 0 \right)^T
\end{equation*}
we can rewrite \eqref{r_s} as
\begin{align*} \nonumber
	r_u & = \rbar + e^{-\kappa (u-t)}(r_t - \rbar) + \sigma_r \int_t^u e^{-\kappa (u-v)} dW^r_v  \\ 
	& = \epsilon_0 + \epsilon_1^T \left(\xibar + e^{-\boldGamma (u-t)} \xi_t \right)  + \epsilon_1^T \int_t^u  e^{-\boldGamma (u-v)} \boldxi dW_v 
\end{align*}
which upon insertion into \eqref{Xs_definition} yields
\begin{align*} \nonumber
	X_s = &  \int_t^s  \Big\{ \underbrace{\left[\epsilon_0 + \epsilon_1^T \left(\xibar + e^{-\boldGamma (u-t)} \xi_t \right)  + \epsilon_1^T \int_t^u  e^{-\boldGamma (u-v)} 
	\boldxi dW_v\right]}_{r_u} +  f_u^T \xi_u \Big\} du \\ \nonumber
	& \qquad \qquad - \frac{1}{2} \int_t^s f_u^T \boldsymbol{C} f_u du + \int_t^s f_u^T dW_u \\ \nonumber
	= & \int_t^s  \left\{ \epsilon_0 +
	\left(\epsilon_1 +  f_u \right)^T  \left(\xibar +  e^{-\boldGamma (u-t)}\xi_t(t) \right) - \frac{1}{2} f_u^T \boldsymbol{C} f_u 
	\right\} du  \\
	& \qquad \qquad + \int_t^s  \left[ (\epsilon_1 + f_u)^T \int_t^u  e^{-\boldGamma (u-v)} \boldxi dW_v \right] du
	+ \int_t^s f_u^T dW_u.
\end{align*}
By interchanging the order of integration of the double integral
\begin{equation*}
	\int_t^s  \left[\int_t^u   (\epsilon_1 + f_u)^T  e^{-\boldGamma (u-v)} \boldxi dW_v \right] du = 
	\int_t^s   \left[ \int_u^s  (\epsilon_1 + f_v)^T  e^{-\boldGamma (v-u)} \boldxi dv \right] dW_u
\end{equation*}
we find
\begin{multline} \nonumber
	X_s = \int_t^s  \left\{ \epsilon_0 +
	\left(\epsilon_1 +  f_u \right)^T  \left(\xibar +  e^{-\boldGamma (u-t)}\xi_t(t) \right) - \frac{1}{2} f_u^T \boldsymbol{C} f_u 
	\right\} du  \\
	+ \int_t^s \left\{ f_u^T + \int_{u}^s (\epsilon_1 +  f_v)^T  e^{-\boldGamma (v-u)} \boldxi dv \right\} dW_u.
\end{multline}
It is well known, \cite{bjork04}, that stochastic integrals of deterministic integrands are Normally distributed, hence, it follows that $\log (V_s/V_t) = X_s$ is normally distributed with mean
\begin{equation*} \label{proof_prop_meanvar_mean}
	\E \log (V_s/V_t) = \int_t^s  \left\{ \epsilon_0 +  
	\Big( \epsilon_1 + f_u \Big)^T  \left(\xibar +  e^{-\boldGamma (u-t)}\xi_t(t) \right) - \frac{1}{2} f_u^T \boldsymbol{C}  f_u 
	\right\} du 
\end{equation*}
and variance
\begin{equation*}  \label{proof_prop_meanvar_variance}
	\V \log(V_s / V_t) = \int_t^s h_u^T \boldsymbol{C} h_u du
\end{equation*}
where
\begin{equation*}
	h_u = f_u + \boldxi \int_u^s  e^{-\boldGamma (v-u)}  (\epsilon_1 +  f_v)  dv
\end{equation*}
where it was utilized that $\boldGamma$ and $\boldxi$ commute. $\square$

%% file: thm_nominal_euler.tex
\section{Proof of Theorem \ref{theorem_nominal_Euler}} \label{proof_theorem_nominal_Euler}
{\em Proof}. All vector and matrix manipulations follow the Jacobian - or numerator - formulation.
First, notice  that from the definition of $p_u$, \eqref{Idefinition}, it follows that
\begin{equation} \label{nominal_dpdx}
	\frac{\partial p}{\partial y} = (\boldGamma l_u )^T \quad \mbox{and} \quad \frac{\partial p}{\partial \dot{y}} = -l_u^T
\end{equation}
where it was utilized that
\begin{equation*} 
	l_u  = \xibar + e^{-\boldGamma (u-t)}\xi_t  - \boldsymbol{C} (\boldGamma y_u - \dot{y}_u)
\end{equation*}
satisfies the relation
\begin{equation*}
	\boldGamma l_u + \dot{l}_u  = \boldsymbol{C} \ddot{y}_u  + (\boldGamma \boldC - \boldC \boldGamma ) \dot{y}_u
	-  \boldGamma\boldsymbol{C} \boldGamma y_u  + \boldGamma\xibar.
\end{equation*}

Similarly, from the definition of $q_u$, \eqref{Jdefinition}, it follows that
\begin{equation} \label{nominal_dqdx}
	\frac{\partial q}{\partial y} = 2 \big( \boldGamma_\xi \boldsymbol{C} h_u \big)^T  \quad \mbox{and} \quad 
	\frac{\partial q}{\partial \dot{y}}  = - 2 (\boldsymbol{C} h_u)^T
\end{equation}
where $\boldGamma_\xi = \boldGamma + \boldxi$ and it was utilized that $h_u$, \eqref{thm_nominal_hu_definition}, satisfies the relation
\begin{align} \nonumber
	\boldGamma_\xi \boldC h_u + \boldC  \dot{h}_u  & = -\boldC \ddot{y}_u + (\boldC \boldGamma_\xi - \boldGamma_\xi \boldC) \dot{y}_u + \boldGamma_\xi \boldC \boldGamma_\xi y_u \\ \nonumber
	& \qquad\qquad  - \boldC \boldxi \epsilon_1 
	+ (\boldGamma_\xi \boldC  + \boldC \boldGamma) \psi_\kappa(s-u) \boldxi \epsilon_1.
\end{align}
 
With these intermediaries, the Euler-Lagrange\ equation, \eqref{thm-nominal-general-Euler-lagrange}, becomes
\begin{multline} 
	\Big\{ \boldGamma l_u - (-\dot{l}_u) \Big\} 
	 -\nu \Big\{ \boldGamma_\xi \boldsymbol{C} h_u  - (- \boldsymbol{C} \dot{h}_u) \Big\} = \\
	\Big\{ (1 +\nu) \boldC \ddot{y}_u + 
	\big[ (\boldGamma\boldC - \boldC\boldGamma)  +\nu (\boldGamma_\xi\boldC - \boldC\boldGamma_\xi) \big] \dot{y}_u  + \boldGamma\xibar  \\
	- \big[\boldGamma\boldC\boldGamma + \nu \boldGamma_\xi\boldC\boldGamma_\xi  \big] y_u 
	 +\nu \Big[ \boldC - (\boldGamma_\xi \boldC + \boldC \boldGamma)\psi_\kappa(s-u)\ \Big] \eta^r \Big\}= 0
\end{multline}
where $\eta^r = \boldxi\epsilon_1$, hence, $y_u$ must satisfy the inhomogeneous second-order differential equation
\begin{equation*}
	(1 +\nu) \Big[ \boldC\ddot{y}_u + \boldB \dot{y}_u - \boldA y_u \Big] = g_u
\end{equation*}
where
\begin{align} \nonumber
	(1 +\nu) \boldA & = \boldGamma\boldC\boldGamma  +\nu \boldGamma_\xi\boldC\boldGamma_\xi , \\ \nonumber
	(1 +\nu) \boldB &=  (\boldGamma\boldC - \boldC\boldGamma)  +\nu (\boldGamma_\xi\boldC - \boldC\boldGamma_\xi), \\ \nonumber
	g_u & =  -\nu \left[ \boldC - (\boldGamma_\xi \boldC  + \boldC \boldGamma) \psi_\kappa(s-u) \right] \eta^r - \boldGamma\xibar.
\end{align}
or - by Proposition \ref{prop_xi_definition} - in the primary parameterization:
\begin{equation*} 
	\boldA  = \begin{Bmatrix}
		\gamma_r^2 & 0 \\ 
		0 & \gamma_S^2
	\end{Bmatrix}
	+ \rho \begin{Bmatrix}
		0 & a_\nu  \\
		a_\nu & 0
	\end{Bmatrix} \qquad
	\boldB  = \rho \begin{Bmatrix}
		0 & b_{\nu} \\
		-b_{\nu} & 0
	\end{Bmatrix} 
\end{equation*}
with
\begin{align} \nonumber
	&\gamma_r^2  =  \frac{\kappa^2  +\nu a^2 }{1 +\nu} && \gamma_S^2  = \frac{\alpha^2  +\nu (\alpha')^2}{1 +\nu} \\ \nonumber
	&a_\nu  = \frac{\alpha\kappa  +\nu a \alpha'}{1 +\nu} && b_\nu  = \frac{(\kappa-\alpha) +\nu(a-\alpha')}{1 +\nu} 
\end{align}
where $\alpha' = \alpha - \sigma_x/\sigma_S$ and
\begin{equation*}
	g_u  =  -\Big[ 
	\begin{pmatrix}
		\kappa\xibar^r \\ \alpha \xibar^S
	\end{pmatrix}
	 +\nu\sigma_r
	\begin{pmatrix}		
		1 - (\kappa+a) \psi_\kappa(s-u) \\ \rho [1-(\kappa+\alpha')\psi_\kappa(s-u)]
	\end{pmatrix}\Big].
\end{equation*}
Furthermore, combining \eqref{thm-nominal_lower_boundary_condition}, \eqref{nominal_dpdx}, and \eqref{nominal_dqdx} it follows that $y_u$ satisfies the lower boundary condition 
\begin{multline*}
	\left( l_u  -\nu \boldC h_u\right)^T \Big|_{u = t} = \\
	  \big( \xibar + e^{-\boldGamma (u-t)}\xi_t  + \boldsymbol{C} (\boldGamma y_u - \dot{y}_u)  +\nu \boldC \{  \boldGamma_\xi y_u - \dot{y}_u + \psi_\kappa(s-u) \eta^r \} \big)^T \Big|_{u = t,s} = 0,
\end{multline*}
and in combination with the upper boundary condition, \eqref{thm-nominal_upper_boundary_condition}, the boundary conditions become
\begin{align*}
	& b_0 + \big( b_u - [ \boldGamma  +\nu \boldGamma_\xi ] y_u + (1 +\nu)\dot{y}_u \big) \Big|_{u = t} = 0 & \text{(lower boundary)}\\
	& y_u |_{u=s} = 0 & \text{(upper boundary)}
\end{align*}
where
\begin{align*} \nonumber
	b_0 = & \boldC^{-1} \xibar, \\
	b_u = & \boldC^{-1} e^{-\boldGamma(u - t)} \xi_t  -\nu \psi_\kappa(s-u)\eta^r.
\end{align*}
$\square$

%% file: cor_latent_roots.tex
\section{Latent Roots}\label{latent_roots}

\begin{cor}\label{cor:corollary_latent_roots}
	{\bf Latent Roots}. Given the assumptions of Theorem \ref{theorem_nominal_Euler} then the Lambda-matrix $\boldM_\lambda$, \eqref{lambda_matrix}, is given by
	\begin{equation}\label{corollar_nominal_solution_latent_roots_lambda_matrix}
		\boldM_\lambda = \begin{Bmatrix}
			\lambda^2 - \gamma_r^2 & \rho[(\lambda^2 - a_{\nu}) + b_{\nu} \lambda] \\
			\rho[(\lambda^2 - a_{\nu}) - b_{\nu} \lambda] & \lambda^2 - \gamma_S^2
		\end{Bmatrix}
	\end{equation}
	with squared latent roots $\lambda_1^2, \lambda_2^2$. Let the discriminant, $D$, be given by
	\begin{equation*}
		D = (1-\rho^2)\big[\gamma_r^2 - \gamma_S^2\big]^2 + \rho^2 \big[ \gamma_r^2 + \gamma_S^2 - (2a_{\nu}+b_{\nu}^2) \big]^2 - \rho^2 (1-\rho^2) (4a_{\nu} +b_{\nu}^2)b_{\nu}^2
	\end{equation*}	\newline
	then (a) iff $D>0$, the squared latent roots are {\em real, positive, and distinct} and are given by
	\begin{equation*} \label{proof_nominal_solution_latent_roots}
		\lambda_1^2 = \frac{ \gamma_r^2+\gamma_S^2-\rho^2(2a_{\nu} + b_{\nu}^2)   - \sqrt{D}}{2(1-\rho^2)}, \quad
		\lambda_2^2 = \frac{ \gamma_r^2+\gamma_S^2-\rho^2(2a_{\nu} + b_{\nu}^2)  + \sqrt{D}}{2(1-\rho^2)}
	\end{equation*}
and the solvents, $\boldS_1, \boldS_2 \in \R^2$ are given by
\begin{equation*} \label{theorem_nominal_solution_solvents}
	\qquad \boldS_1 = \boldQ_1 \boldLambda \boldQ_1^{-1}, \qquad \boldS_2 = \boldQ_2 (-\boldLambda) \boldQ_2^{-1} 
\end{equation*}
where $\boldLambda = {\rm diag}(\lambda_1, \lambda_2)$ and $\boldQ_1, \boldQ_2 \in \R^{2\times 2}$, 
$\boldQ_1 = \{r_1 ; r_2\}$ and $\boldQ_2 = \{r_3 ; r_4\}$ , with column right-latent vectors, $r_i$, where $r_i$ is the larger of
\begin{equation*}
	\bar{r}_i = \begin{pmatrix}
		\rho [ \lambda_i^2 - a_{\nu} - b_{\nu} \lambda_i ] \\ \gamma_r^2 - \lambda_i^2 
	\end{pmatrix} 
	\qquad	
	\underbar{r}_i = \begin{pmatrix}
		\gamma_S^2 - \lambda_i^2 \\ \rho [ \lambda_i^2 - a_{\nu} + b_{\nu} \lambda_i ]
	\end{pmatrix}
\end{equation*}
with respect to the Euclidian norm for $\lambda_i = (\lambda_1, \lambda_2,-\lambda_1,-\lambda_2)$, respectively.
\newline
	and (b) iff $D<0$, the latent roots are {\em complex, distinct, and each others complex conjugate} and are given by
	\begin{equation*} \label{proof_nominal_solution_complex_latent_roots}
		\lambda_1^2 = (\lambda_2^2)^* = \frac{\gamma_r^2+\gamma_S^2-\rho^2(2a_{\nu} + b_{\nu}^2)  -i \sqrt{-D}}{2(1-\rho^2)}
	\end{equation*}
	where $(\cdot)^*$ denotes the complex conjugate and the solvents, $\boldS_1,\boldS_2$, are {\em real} and given by
	\begin{equation*}\label{lemma_nominal_harmonic_solvents}
		\boldS_i = \frac{1}{{\rm Im}(r_{i1}r_{i2}^*)} {\rm Im}\begin{Bmatrix}
			\lambda r_{i1}r_{i2}^* & \lambda^* |r_{i1}|^2 \\
			\lambda |r_{i}|^2 & \lambda^* r_{i1} r_{i2}^* 
		\end{Bmatrix},
	\end{equation*}
	$\lambda$ is the principal (complex) root of
	\begin{equation*}
		\lambda^2 = \frac{\gamma_r^2+\gamma_S^2-\rho^2(2a_{\nu} + b_{\nu}^2)  -i \sqrt{-D}}{2(1-\rho^2)},
	\end{equation*}
	the latent vectors $r_1,r_2 \in \C^2$ are given by
	\begin{equation*}
		r_1 = \begin{pmatrix}
			\rho [\lambda^2 - a_{\nu} + b_{\nu} \lambda] \\ \gamma_r^2 - \lambda^2.
		\end{pmatrix}, \qquad
		r_2 = \begin{pmatrix}
			\rho [\lambda^2 - a_{\nu} - b_{\nu} \lambda] \\ \gamma_r^2 - \lambda^2,
		\end{pmatrix},
	\end{equation*}
	respectively.
		
	\begin{proof}$D>0$: 
		It follows from \eqref{corollar_nominal_solution_latent_roots_lambda_matrix} that the determinant is a quadratic polynomial in $\lambda^2$
		\begin{equation*}
			\det \boldM_\lambda = \underbrace{(1-\rho^2)}_A \lambda^4 - \underbrace{\Big[(\gamma_r^2 + \gamma_S^2) - \rho^2(2a_{\nu} + b_{\nu}^2)\Big]}_B \lambda^2 + \underbrace{\big[\gamma_r^2\gamma_S^2 - \rho^2 a_{\nu}^2\big]}_C
		\end{equation*}
		where all coefficients $A,B,$ and $C$ are positive:
		\begin{itemize}
		\item {\em Positivity of $A$} follows directly from the assumption $\rho^2<1$.\newline	
		\item {\em Positivity of $B$} holds, if $\gamma_r^2 + \gamma_S^2 \geq 2a_{\nu} + b_{\nu}^2$. Upon multiplication by $(1 +\nu )^2$ we find
		\begin{align*}
			& (1 +\nu )^2 \big[\gamma_r^2 + \gamma_S^2 - \rho^2(2a_{\nu} - b_{\nu}^2) \big] \geq 0 && \Leftrightarrow \\
			&  +\nu  \Big((\kappa-\alpha) - (a-\alpha')\Big)^2 \geq 0
		\end{align*}
		which holds for all parameter choices since $\nu > 0$.\newline
		\item {\em Positivity of $C$} holds, if $\gamma_r^2\gamma_S^2 > \rho^2 a_{\nu}^2$ since $b_\nu$ is positive. It follows upon multiplication by $(1 +\nu )^2$ that
		\begin{align*}
			& \big[\kappa^2 +\nu  a^2)\big]\big[\alpha^2 +\nu (\alpha')^2\big] - \rho^2\big[\alpha\kappa +\nu  a\alpha'\big]^2 > 0 &&\Leftrightarrow \\
			& (1-\rho^2)(\alpha^2\kappa^2 + \nu^2 a^2(\alpha')^2) 
			 +\nu (\kappa\alpha' - a \alpha)^2 > 0	
		\end{align*}
		which also holds for all parameter choices.
		\end{itemize}
		Since $B$ is positive, the vertex is positive, hence, the larger root is positive. Furthermore, since $A$ is positive, the parabola opens upwards, and since $C$ (the intersection) is also positive, the smaller root is positive too. Finally, since $D>0$ the roots are distinct.
		
		The lambda matrix is degenerate at the latent roots, hence, the top and bottom row of $\boldM_\lambda$ become proportional. The latent (right) vectors, $r_i$, solve
		\begin{equation}\label{cor_roots_eigenvalue_equation}
			\boldM_{\lambda_i} r_i = 0 
		\end{equation}
		hence, the right latent vector is given by
		\begin{equation*}\label{proof_nominal_solution_latent_vectors}
			r_i = \begin{pmatrix}
				\rho [ \lambda_i^2 - a_{\nu} - b_{\nu} \lambda_i ] \\ \gamma_r^2 - \lambda_i^2
			\end{pmatrix} 
		\end{equation*}
		or in case this is a zero-vector
		\begin{equation*}\label{proof_nominal_solution_latent_vectors_alternative}
			r_i = \begin{pmatrix}
				\gamma_S^2 - \lambda_i^2 \\ \rho [ \lambda_i^2 - a_{\nu} + b_{\nu} \lambda_i ]
			\end{pmatrix}.
		\end{equation*}
		Furthermore, right latent vectors of the pair $\lambda_1$ and $\lambda_2$ are linearly independent.
		\newline
		$D<0$: 
			The latent roots are complex and given by
		\begin{equation*}
			\lambda_1^2 = (\lambda_2^2)^* = \frac{\gamma_r^2+\gamma_S^2-\rho^2(2a_{\nu} + b_{\nu}^2)  -i \sqrt{-D}}{2(1-\rho^2)}
		\end{equation*}
		hence, there are four distinct (complex) roots, $\bar{\gamma}_i = \{\bar{\gamma}, \bar{\gamma}^*, -\bar{\gamma}, -\bar{\gamma}^* \}$ where $\bar{\gamma}$ is the principal square root of $\lambda_1^2$.
		
		Since the latent roots are complex, the latent vectors are given by \eqref{cor_roots_eigenvalue_equation} and from the ordering of the latent roots,$\lambda_i$, $\boldQ_1$ and $\boldQ_2$ are given by
		\begin{align*}
			\boldQ_1 & = \{r_1;r_1^*\}, &&r_1 = \begin{pmatrix}
				\rho [\gamma^2 - a_{\nu} + b_{\nu} \gamma] \\ \gamma_r^2 - \gamma^2.
			\end{pmatrix}\\
			\boldQ_2 & = \{r_2;r_2^*\}, &&r_2 = \begin{pmatrix}
				\rho [\gamma^2 - a_{\nu} - b_{\nu} \gamma] \\ \gamma_r^2 - \gamma^2.
			\end{pmatrix}.
		\end{align*}
		For either $\boldQ_1$ or $\boldQ_2$ we write with a minor abuse of notation
		\begin{equation*} 
			\boldQ = \begin{Bmatrix}
				r_{1} 	& r_{1}^* \\ r_{2} & r_{2}*
			\end{Bmatrix} \Rightarrow
			\boldQ^{-1} = \frac{-i}{2{\rm Im}{(r_{1}r_{2}*)}} \begin{Bmatrix}
				r_{2}* 	& -r_{1}^* \\ -r_{2} & r_{1}
			\end{Bmatrix}
		\end{equation*}
		and therefore
		\begin{equation*} \label{proof_nominal_solution_harmonic_solvents}
			\boldS = \boldQ \boldLambda \boldQ^{-1} 
			= \frac{1}{{\rm Im}(r_{1}r_{2}*)} {\rm Im}\begin{Bmatrix}
				\gamma r_1r_2^* & \gamma^* |r_1|^2 \\
				\gamma |r_2|^2 & \gamma^* r_1 r_2^* 
			\end{Bmatrix},
		\end{equation*}
		that is, the {\em solvents} $\boldS_1,\boldS_2$ are real. $\square$	
	\end{proof}
\end{cor}

%% file: thm_nominal_solution.tex
\section{Proof of Theorem \ref{theorem_nominal_solution}} \label{proof_nominal_solution}
Writing the solution
\begin{equation*}
	y_u = y_u^p + y_u^h
\end{equation*}
as the sum of the particular, $y_u^p$ and homogeneous, $y_u^h$ parts, respectively, by Theorem \ref{thm_spectral_problem} the homogeneous solution is given by 
\begin{equation} \label{homogeneous_solution}
	y_u^h = e^{\boldS_1(s-u)} q_1 + e^{\boldS_2(s-u)} q_2
\end{equation}
where $q_1, q_2 \in \R^2$ are arbitrary integration constants. It will become clear below, that $q_1, q_2$ depend explicitly on the initial time, $t$, but for the purposes of this proof we will suppress this dependence. 

For the particular solution, we assume the following form
\begin{equation} \label{particular_solution}
	y^p_u  = k_1 + \psi_\kappa(s-u) k_2
\end{equation}
where $k_1, k_2 \in \R^2$ are constants. Upon insertion into \eqref{thm_nominal_equation} and matching factors of $\psi_\kappa(s-u)$, we find
\begin{subequations}
	\begin{numcases}{} \label{particular_k2}
		(1+\nu) \Big[ \kappa^2 \boldC + \kappa \boldB  - \boldA  \Big] k_2  =  \nu  \sigma_r 
		\begin{pmatrix}
			a+\kappa \\ \rho(\alpha' + \kappa)
		\end{pmatrix} 
		\\   \label{particular_k1}
		(1+\nu)\Big[ (\kappa\boldC + \boldB) k_2 + \boldA k_1 \Big] = 
		\begin{pmatrix}
			\kappa \xibar^r  \\  \alpha\xibar^S 
		\end{pmatrix}
		 +\nu \sigma_r
		\begin{pmatrix}
			1 \\  \rho
		\end{pmatrix}
	\end{numcases}
\end{subequations}
Rewriting the left-hand side of \eqref{particular_k2}
\begin{align*} \nonumber
	(1+\nu) & \big\{ \kappa^2 \boldC + \kappa \boldB  - \boldA  \big\} k_2  \\ \nonumber & = 
	\begin{Bmatrix}
		\nu(a+\kappa)(\kappa-a) & \rho[2\kappa(k-\alpha)  +\nu (\kappa-\alpha')(\kappa+a)] \\
		\nu\rho(\kappa+\alpha')(\kappa-a) & (1 +\nu )(\kappa^2 - \gamma_S^2)
	\end{Bmatrix} k_2 
\end{align*}
it is clear that the first column is proportional to the right-hand side of \eqref{particular_k2}, hence,
\begin{equation*} \label{proof_nominal_solution_k2}
	k_2 = \Big(
	\frac{\sigma_r}{\kappa-a} , 0
	\Big)^T.
\end{equation*}
Insertion of $k_2$ into \eqref{particular_k1} yields $k_1$:
\begin{equation*}
	(1 +\nu )\boldA k_1 = \begin{pmatrix}
		\kappa \xibar^r \\ \alpha \xibar^S
	\end{pmatrix}
	+ \frac{\sigma_r}{a-\kappa}
	\begin{pmatrix}
		\kappa  +\nu  a \\ \rho (\alpha  +\nu  \alpha')
	\end{pmatrix}
\end{equation*}
where the inverse of $\boldA$ is given by
\begin{equation*}
	\boldA^{-1}
	=\frac{1}{\gamma_r^2\gamma_S^2 - \rho^2 a_{\nu}^2} 
	\begin{Bmatrix}
		\gamma_S^2 & - \rho a_{\nu} \\ - \rho a_{\nu} & \gamma_r^2
	\end{Bmatrix}.
\end{equation*}

The boundary conditions are stated in Theorem \ref{theorem_nominal_Euler} which together with \eqref{homogeneous_solution} and \eqref{particular_solution} yields the following condition to determine integration constants $q_1$ and $q_2$:
\begin{equation*}
	\begin{cases}
		q_1 + q_2 + k_1 = 0
		\\
		b_0 + b_t 
		- \big[ \boldGamma  +\nu  \boldGamma_\xi  \big] 
		\Big(k_1 + k_2\psi_\kappa(s-t) + e^{\boldS_1(s-t)} q_1 + e^{\boldS_2(s-t)} q_2 \Big)
		\\ +\big[1 +\nu \big]  \Big(  \big[\kappa \psi_\kappa(s-t) - 1)\big] k_2 
		-\boldS_1 e^{\boldS_1(s-t)} q_1 - \boldS_2 e^{\boldS_2(s-t)} q_2 \Big) = 0
	\end{cases}
\end{equation*}
-- or on block matrix form
\begin{equation*} \label{proof_nominal_solution_boundary_conditions}
	\begin{Bmatrix}
		\unitI & \unitI
		\\
		\boldD_1  e^{\boldS_1 (s-t)} & \boldD_2  e^{\boldS_2 (s-t)}
	\end{Bmatrix} 
	\begin{pmatrix}
		q_1 \\ q_2
	\end{pmatrix} 
	= \begin{pmatrix}
		-k_1 \\
		\boldC^{-1} (\xibar + \xi_t) - (\boldGamma k_1 + k_2)  -\nu (\boldGamma_\xi k_1 + k_2)
	\end{pmatrix}.
\end{equation*}
where $\boldD_1 = \boldGamma  +\nu \boldGamma_\xi + (1 +\nu )\boldS_1$ and $\boldD_2 = \boldGamma  +\nu \boldGamma_\xi + (1 +\nu )\boldS_2$.

Finally, by Theorem \ref{theorem_nominal_Euler}, \eqref{homogeneous_solution}, and \eqref{particular_solution} it follows that the optimal factor allocation, $f_u$, is given by
\begin{align*}
	f_u & = \boldGamma y_u - \dot{y}_u  \\ 
	& = \boldGamma \big\{ k_1 + \psi_\kappa(u) k_2 + e^{\boldS_1 (s-u)}q_1 + e^{\boldS_2(s-u)} \big\} \\  
	&  \qquad - (\kappa \psi_\kappa(u) - 1)k_2 + \boldS_1 e^{\boldS_1(s-u)}  q_1 + \boldS_2 e^{\boldS_2(s-u)} q_2 \\
	& = (\boldGamma k_1 + k_2) + (\boldGamma + \boldS_1) e^{\boldS_1 (s-u)} q_1 + (\boldGamma + \boldS_2) e^{\boldS_2 (s-u)} q_2.
\end{align*}
$\square$

%% file: lemma_meanvar.tex
The horizon mean and horizon variance is stated in the following Lemma:
\begin{lem} \label{lemma_meanvar}\label{lem:meanvar}
    Given Proposition \ref{theorem_nominal_Euler} and Theorem \ref{theorem_nominal_solution}, the horizon mean is given by
    \begin{align*}
        \E \log &(V_s/V_t)  = \\
        & m^0 + \sum_{i\in \{0,1,2\}} \Big[\xibar^T (\boldGamma + \boldS_i)\boldPsi_{-\boldS_i}(s-t) + 
        \xi_t^T (\boldS_i \boldPsi_{-\boldS_i}(s-t) + \boldGamma \boldPsi_{\boldGamma}(s-t))
        \Big]  q_{it} 
        \\& \qquad \qquad \qquad - \frac{1}{2} \sum_{ij\in \{0,1,2\}}  q_{it}^T \big(\boldQ_i^{-1}\big)^T \Big[\boldH^{ij} \odot \boldK_{\boldLambda_i,\boldLambda_j}(s-t)
                    \Big] \, \boldQ_j^{-1} q_{jt}
    \end{align*}
    with the convention
    \begin{equation} \label{lem:meanvar:S0}
        \boldLambda_i= \{\boldZero, \boldLambda, -\boldLambda \}  \qquad q_{0t} = (k_1 + k_2/\kappa), \qquad \boldS_0 = \boldZero, \qquad \boldQ_0 = \boldQ_0^{-1} = \unitI
    \end{equation}
    for $i\in \{0,1,2\}$ where the 'cash only' horizon mean, $m^0$, is given by
    \begin{equation*}
        m^0 = \rbar(s-t) + \psi_{\kappa}(s-t)(r_t-\rbar)
    \end{equation*}
    and
    \begin{align} \nonumber  
        \boldH^{ij} & = \boldQ_i^T
        (\boldGamma + \boldS_i)^T \boldC (\boldGamma + \boldS_j)\boldQ_j \\ \label{lem:meanvar:K}
        \Big(\boldK_{\bar{\boldD}\underaccent{\bar}{\boldD}}(s-t)\Big)_{lm} & = 
        \begin{cases}
            (s-t) & \bar{d}_l + \underaccent{\bar}{d}_m = 0 \\
            \frac{e^{(\bar{d}_l + \underaccent{\bar}{d}_m) (s-t)}-1}{\bar{d}_l+\underaccent{\bar}{d}_m} & {\rm else}
        \end{cases}  
    \end{align}
    for $l,m\in \{1,2 \}$ where $\bar{\boldD}, \underaccent{\bar}{\boldD} \in \R^2$ are diagonal matrices and $\bar{d}_{1},\underaccent{\bar}{d}_{1} $ are upper diagonal and $\bar{d}_{2},\underaccent{\bar}{d}_{2}$ lower diagonal elements of $\bar{\boldD},\underaccent{\bar}{\boldD}$, respectively,  $\odot$ is the Hadamard product, and 
    \begin{equation}\label{lem:meanvar:PSI}
        \boldPsi_{\boldM}(s-t) = 
        \begin{cases}
            (s-t)\unitI & \text{for } \boldM = \boldZero \\
            \boldM^{-1} \left(\unitI - e^{-\boldM(s-t)}\right) & \text{$\boldM$ is full rank }
        \end{cases}
    \end{equation}
    Furthermore, the conditional horizon variance is given by
    \begin{align*}
        \V \log&(V_s / V_t)  = k_2^T k_2 (s-t) + 2 k_2^T \boldC \sum_{i\in\{0,1,2\}} \Big[
            (\boldGamma + \boldxi + \boldS_i)\boldPsi_{-\boldS_i}(s-t)
            - \boldxi \boldPsi_{\boldGamma}(s-t)
            \Big] \tilde{q}_{it} \\
            &+  \sum_{i,j \in \{0,1,2\}} \Big\{ \tilde{q}^T_{it}
            \big(\boldQ_i^{-1}\big)^T \left[\boldH^{ij}_{\boldxi} \odot \boldK_{\boldLambda_i,\boldLambda_j}\right]\, \boldQ_j^{-1} \\
            &  \qquad  \qquad - (\boldQ^{-1}_i)^T \left[\boldG^{ij} \odot \boldK_{\boldLambda_i,-\boldGamma}\right] 
           - \left[ \boldK_{-\boldGamma,\boldLambda_j} \odot (\boldG^{ji})^T \right]\, \boldQ_j^{-1}  + \boldC_{\boldxi} \odot \boldK_{-\boldGamma,-\boldGamma}  \Big\}\tilde{q}_{jt}
    \end{align*}
    where $\tilde{q}_{0t}=k_1$ and $\tilde{q}_{it}=q_{it}$ for $i\in(1,2)$ and
    \begin{align*}
        \boldH^{ij}_{\boldxi} &= \boldQ^T_i  (\boldGamma + \boldxi + \boldS_i)^T  \boldC 
        (\boldGamma + \boldxi + \boldS_i) \boldQ_j \\
        \boldG^{ij} &= (\boldGamma + \boldxi + \boldS_i)^T \boldC \boldxi \\
        \boldC_{\boldxi} &= \boldxi  \boldC  \boldxi
    \end{align*}

\begin{proof}
    The conditional mean is given by \eqref{nominal_mean}. First, the 'cash only' conditional mean, $m^0$, for $f_u\equiv 0$ is given by
    \begin{equation*}
        m^0 = \int_t^s \big\{\epsilon_0 + \epsilon_1^T (\xibar + e^{-\boldGamma(u-t)}\xi_t)\} = \rbar(s-t) + \psi_{\kappa}(s-t)(r_t-\rbar)
    \end{equation*}
    where $\psi_{\kappa}(\cdot)$ is given by \eqref{psi} and we have recovered \eqref{capmkt_cash_mean}. 
   
    From \eqref{thm:fu} and using the convention \eqref{lem:meanvar:S0} we can rewrite $f_u$ as
    \begin{equation} \label{lem:meanvar:fu}
        f_u = \sum_{i\in \{0,1,2\} } (\boldGamma + \boldS_i) e^{\boldS_i (s-u)} q_{it},
    \end{equation}
    hence, terms proportional to $f_u$ in the first integral of \eqref{nominal_mean} are 
    \begin{equation*}
        \int_t^s \Big[\xibar + e^{-\boldGamma (u-t)} \xi_t \Big]^T (\boldGamma + \boldS_i) e^{\boldS_i (s-u)} \,du \quad i \in \{ 0,1,2 \}
    \end{equation*}
    where for terms proportional to $\xibar^T$:
    \begin{equation*}
        (\boldGamma + \boldS_i)\int_t^s e^{\boldS_i (s-u)} du  = 
        (\boldGamma + \boldS_i)\boldPsi_{-\boldS_i}(s-t)
    \end{equation*}
    using \eqref{lem:meanvar:PSI}. Next, consider the integral
    \begin{equation*}
        \int_t^s e^{-\boldGamma (u-t)} e^{\boldS_i (s-u)} du =
        -\Big[
            \boldGamma^{-1} e^{-\boldGamma(u-t)} e^{\boldS_i(s-u)}
        \Big]_{u=t}^{u=s} - \boldGamma^{-1} \int_t^s e^{-\boldGamma (u-t)}  e^{\boldS_i (s-u)} du \,\boldS_i
    \end{equation*}
    by integration of parts, hence, terms proportional to $\xi_t^T$ are given by
    \begin{align}\label{lem:meanvar:special_integral}
        \int_t^s e^{-\boldGamma (u-t)} (\boldGamma + \boldS_i) e^{\boldS_i (s-u)} du & = 
        \Big[
            e^{\boldS_i(s-t)} - e^{-\boldGamma(s-t)} 
        \Big] .
        \end{align}
    For the quadratic terms we find
    \begin{align*}    
        \int_t^s & \Big(e^{\boldS_i (s-u)}\Big)^T
        (\boldGamma + \boldS_i)^T \boldC (\boldGamma + \boldS_j) e^{\boldS_j (s-u)}du \\
        &= \big(\boldQ_i^{-1}\big)^T \int_t^s e^{\boldLambda_i (s-u)} \underbrace{ \boldQ_i^T
        (\boldGamma + \boldS_i)^T \boldC (\boldGamma + \boldS_j)\boldQ_j}_{\boldH^{ij}}  e^{\boldLambda_j (s-u)} du \, \boldQ_j^{-1} \\
        &= \big(\boldQ_i^{-1}\big)^T \int_t^s
        \begin{Bmatrix}
            h^{ij}_{11} e^{(\lambda^{(i)}_1+\lambda^{(j)}_1) (s-u)} &  h^{ij}_{12} e^{(\lambda^{(i)}_1+\lambda^{(j)}_2)(s-u)} \\
            h^{ij}_{21} e^{(\lambda^{(i)}_2+\lambda^{(j)}_1) (s-u)} &  h^{ij}_{22} e^{(\lambda^{(i)}_2+\lambda^{(j)}_2) (s-u)}
        \end{Bmatrix} du \, \boldQ_j^{-1} \\
        =& \big(\boldQ_i^{-1}\big)^T \left[\boldH^{ij} \odot \boldK_{\boldLambda_i,\boldLambda_j}
        \right] \, \boldQ_j^{-1}
    \end{align*}
    where $\lambda^{(i)}_{1}$ is the upper diagonal element and $\lambda^{(i)}_2$ the lower diagonal element of $\Lambda_i$, respectively, $h^{ij}_{lm}$ are the individual elements of $\boldH^{ij}$, $\odot$ is the Hadamard matrix product, and $\boldK$ is given by \eqref{lem:meanvar:K} from which the result follows.
 
    {\em Horizon Variance.} For the horizon variance, we first combine \eqref{h_nominal_definition}, \eqref{lem:meanvar:fu}, and \eqref{lem:meanvar:special_integral} to rewrite $h_u$ as
    \begin{align}\nonumber
        h_u  & = \sum_{i\in(0,1,2)} \Big[ (\boldGamma + \boldS_i) e^{\boldS_i(s-u)}
        + \boldxi \big( e^{\boldS_i(s-u)} - e^{-\boldGamma(s-u)}  \big)  \Big]  q_{it}
        + \boldxi \int_u^s  e^{-\boldGamma (v-u)}   dv\, \epsilon_1 \\\label{lem:meanvar:h_u}
         &=k_2 + \sum_{i\in(0,1,2)} \Big[ (\boldGamma + \boldxi + \boldS_i) e^{\boldS_i(s-u)}
        - \boldxi  e^{-\boldGamma(s-u)}  \Big]  \tilde{q}_{it}
    \end{align}
    where $\tilde{q}_{0t}=k_1$ and $\tilde{q}_{it}=q_{it}$ for $i\in(1,2)$ utilizing that for the term $i\equiv 0$ 
    \begin{gather*}
        \Big[ (\boldGamma + \boldS_0) e^{\boldS_0(s-u)}
        + \boldxi \big( e^{\boldS_0(s-u)} - e^{-\boldGamma(s-u)}  \big)  \Big]  q_{0t}
        + \boldxi \int_u^s  e^{-\boldGamma (v-u)}   dv\, \epsilon_1 = \\
        k_2 + \Big[\boldGamma +  \boldxi \big( \unitI - e^{-\boldGamma(s-u)} \big)\Big]k_1
    \end{gather*}
    and $k_2 = -\epsilon_1$.

    For the horizon variance, first consider the quadratic terms in $\tilde{q}_{it},\tilde{q}_{jt}$ inserting \eqref{lem:meanvar:h_u} into \eqref{nominal_variance}:
    \begin{align*}
        \int_t^s&
        \Big[ (\boldGamma + \boldxi + \boldS_i) e^{\boldS_i(s-u)}
        - \boldxi  e^{-\boldGamma(s-u)}  \Big]^T  \boldC 
        \Big[ (\boldGamma + \boldxi + \boldS_j) e^{\boldS_j(s-u)}
        - \boldxi  e^{-\boldGamma(s-u)}  \Big] du  \\
        &=(\boldQ^{-1}_i)^T \int_t^s  e^{\boldLambda_i(s-u)} \underbrace{\boldQ^T_i  (\boldGamma + \boldxi + \boldS_i)^T  \boldC 
            (\boldGamma + \boldxi + \boldS_j) \boldQ_j}_{\boldH^{ij}_{\boldxi}} e^{\boldLambda_j(s-u)}  du \, \boldQ^{-1}_j  \\
            & \qquad - (\boldQ^{-1}_i)^T  \int_t^s e^{\boldLambda_i(s-u)} \underbrace{ \boldQ^T_i  (\boldGamma + \boldxi + \boldS_i)^T \boldC \boldxi   }_{\boldG^{ij}}  e^{-\boldGamma(s-u)} du \,           \\
            & \qquad - \int_t^s  e^{-\boldGamma(s-u)} \underbrace{ \boldxi  \boldC  (\boldGamma + \boldxi + \boldS_j) \boldQ_j}_{(\boldG^{ji})^T} e^{\boldLambda_j(s-u)} du \, \boldQ^{-1}_j \\
            & \qquad +  \int_t^s  e^{-\boldGamma(s-u)} \underbrace{\boldxi  \boldC  \boldxi}_{\boldC_{\boldxi}}  e^{-\boldGamma(s-u)} 
            du \\
            & = \big(\boldQ_i^{-1}\big)^T \left[\boldH^{ij}_{\boldxi} \odot \boldK_{\boldLambda_i,\boldLambda_j}\right]\, \boldQ_j^{-1} 
            - (\boldQ^{-1}_i)^T \left[\boldG^{ij} \odot \boldK_{\boldLambda_i,-\boldGamma}\right] 
           \\
            & \qquad - \left[ \boldK_{-\boldGamma,\boldLambda_j} \odot (\boldG^{ji})^T \right]\, \boldQ_j^{-1}  + \boldC_{\boldxi} \odot \boldK_{-\boldGamma,-\boldGamma},
    \end{align*}
    since $\boldGamma$ and $\boldxi$ commute, $k_2 = (\sigma_r/(\kappa-a),0)^T$, and using \eqref{lem:meanvar:K}. For the remaining terms
    \begin{align*}
        \int_t^s h_u^T \boldsymbol{C} &h_u du =  k_2^T k_2 (s-t) \\
        & + 2 k_2^T \boldC \sum_{i\in\{0,1,2\}} \Big[
                (\boldGamma + \boldxi + \boldS_i)\boldPsi_{-\boldS_i}(s-t)
                - \boldxi \boldPsi_{\boldGamma}(s-t)
                \Big] \tilde{q}_{it} + \ldots
    \end{align*} 
    using \eqref{lem:meanvar:h_u} from which the results follows.  $\square$
    \end{proof}
\end{lem}

%% file: cor_infiniteriskaversion.tex
\subsection{Proof of Corollary \ref{cor:infinite_riskaversion}} \label{proof:infinite_riskaversion}
From Theorem \ref{theorem_nominal_Euler} we find in the limit $\nu\rightarrow \infty$ that
\begin{equation*} 
	\gamma_r^2  \rightarrow a^2  \qquad \gamma_S^2  \rightarrow (\alpha')^2 \qquad
	a_\nu  \rightarrow a \alpha' \qquad b_\nu \rightarrow (a-\alpha')
\end{equation*}
and
\begin{equation*}
	\boldA  \rightarrow \begin{Bmatrix}
		a^2 & \rho a \alpha'  \\ 
		\rho a \alpha' & (\alpha')^2
	\end{Bmatrix}
	\qquad
	\boldB \rightarrow \begin{Bmatrix}
		0 & \rho(a-\alpha') \\
		-\rho(a-\alpha') & 0
	\end{Bmatrix} ,
\end{equation*}
hence, the limiting value of the lambda matrix, \eqref{lambda_matrix}, becomes
\begin{equation*}
	\boldM_\lambda \rightarrow \begin{Bmatrix}
		\lambda^2 - a^2 & \rho [\lambda^2 - a \alpha' +(a-\alpha')\lambda]  \\ 
		 \rho [\lambda^2 - a \alpha' - (a-\alpha')\lambda]  & \lambda^2 - (\alpha')^2
	\end{Bmatrix}
\end{equation*}
with limiting latent roots $\lambda^2 = a^2, (\alpha')^2$.

From
\begin{equation*}
	\boldA^{-1}
	\rightarrow \frac{1}{(1-\rho^2)a^2(\alpha')^2} 
	\begin{Bmatrix}
		(\alpha')^2 & - \rho a\alpha' \\ - \rho a\alpha' & a^2
	\end{Bmatrix}
\end{equation*}
it further follows that
\begin{equation*}
	k_1 \rightarrow 
	 \frac{1}{(1-\rho^2)a^2(\alpha')^2} 
	\begin{Bmatrix}
		(\alpha')^2 & - \rho a\alpha' \\ - \rho a\alpha' & a^2
	\end{Bmatrix}
 \frac{\sigma_r}{a-\kappa}
	\begin{pmatrix}
		a \\ \rho \alpha'
	\end{pmatrix}
=  \frac{\sigma_r}{a-\kappa}
	\begin{pmatrix}
	1/a \\ 0
\end{pmatrix},
\end{equation*}
that is, $k_1 \rightarrow -a^{-1}k_2$, hence,
\begin{equation}\label{cor:ksum_is_zero}
	\boldGamma_\xi k_1 + k_2 \rightarrow 0.
\end{equation}

From Theorem \ref{theorem_nominal_solution} we find using \eqref{cor:ksum_is_zero} that for terms proportional to $\nu$, the limiting boundary conditions are
\begin{equation*}
	\begin{Bmatrix}
		\unitI & \unitI
		\\
		( \boldGamma_\xi + \boldS_1 )  e^{\boldS_1 (s-t)} & 0
	\end{Bmatrix} 
	\begin{pmatrix}
		q_1 \\ q_2
	\end{pmatrix} = 
	\begin{pmatrix}
		-k_1 \\
		0
	\end{pmatrix}
\end{equation*}
where it is easy to check that $\boldS_2 = -\boldGamma_\xi$ is a solvent, hence, $\boldD_2 \rightarrow 0$ and
\begin{equation*}
	q_1 = 0 \qquad q_2 = -k_1.
\end{equation*}

Finally, the limiting optimal asset allocation, $f^\infty_u$, is given by insertion in \eqref{thm:fu}
\begin{equation*}
	f^\infty_u = \kappa (-a^{-1})k_2 + k_2 + (\boldGamma - \boldGamma_\xi)e^{-a(s-u)}(-k_1)
	= \begin{pmatrix}
		-\sigma_r \psi_a(s-u) \\ 0
	\end{pmatrix}
\end{equation*}
where $\psi_a(\cdot)$ is given by \eqref{psi}. $\square$